\documentclass[prodmode,acmtecs]{acmsmall} 

\usepackage[ruled]{algorithm2e}

\SetAlFnt{\small}
\SetAlCapFnt{\small}
\SetAlCapNameFnt{\small}
\SetAlCapHSkip{0pt}
\IncMargin{-\parindent}

\acmVolume{1}
\acmNumber{1}
\acmArticle{1}
\acmYear{2016}
\acmMonth{1}

\begin{document}

\markboth{L. CAO}{Data Science: A Comprehensive Overview}

\title{Data Science: A Comprehensive Overview}
\author{LONGBING CAO
\affil{University of Technology Sydney, Australia}}


\begin{abstract}
The twenty-first century has ushered in  the age of big data and data economy, in which \textit{data DNA}, which carries important knowledge, insights and potential, has become an intrinsic constituent of all data-based organisms. An appropriate understanding of data DNA and its organisms relies on the new field of \emph{data science} and its keystone, \emph{analytics}. Although it is widely debated whether big data is only hype and buzz, and data science is still in a very early phase, significant challenges and opportunities are emerging or have been inspired by the research, innovation, business, profession, and education of data science. This paper provides a comprehensive survey and tutorial of the fundamental aspects of data science: the evolution from data analysis to data science, the data science concepts, a big picture of the era of data science, the major challenges and directions in data innovation, the nature of data analytics, new industrialization and service opportunities in the data economy, the profession and competency of data education, and the future of data science. This article is the first in the field to draw a comprehensive big picture, in addition to offering rich observations, lessons and thinking about data science and analytics.
\end{abstract}

%
%
%

%
%






\acmformat{Longbing Cao, 2016. Data Science: A Comprehensive Overview.}

\begin{bottomstuff}
This work is partially supported by the Australian Research Council Discovery Grant, under grant DP130102691.

Author's addresses: L. Cao, Advanced Analytics Institute, University of Technology Sydney, Australia.
\end{bottomstuff}

\maketitle

\section{Introduction}
\label{sec:intro}
We are living in the age of big data, advanced analytics, and data science. The trend of ``big data growth'' \cite{Laney01,CSCbd,Beyer12,Mckinsey11,IDC12}, or ``data deluge'' \cite{Hey03}, has not only triggered tremendous hype and buzz, but more importantly presents enormous challenges which in turn bring incredible innovation and economic opportunities. Big data has attracted intensive and growing attention, initially from giant private data-oriented enterprise and lately from major governmental organizations and academic institutions. Typical examples include large data-centric projects in Google, Facebook and IBM, and strategic initiatives in the United Nations \cite{UNpulse,USNSF}, EU \cite{ECbd14} and China \cite{CNbd}. 

From the disciplinary development perspective, recognition of the significant challenges, opportunities and values of big data is fundamentally reshaping the traditional data-oriented scientific and engineering fields. It is also reshaping those non-traditional data engineering domains such as social science, business and management \cite{yiu12,Labrinidis12,Chen12,Khan14}. This reshaping and paradigm shifting is driven not just by data itself but all other aspects that could be created, transformed and/or adjusted by understanding, exploring and utilizing data. 

The above trend and its potential have triggered new debate about data-intensive scientific discovery as  a new paradigm, the so-called ``fourth science paradigm'', which unifies experiment, theory and computation (corresponding to ``empirical'' or ``experimental'', ``theoretical'' and ``computational'' science) \cite{Gray07,4thparadigm}. Data is regarded as the new Intel Inside \cite{O'reilly05}, or the new oil and strategic asset, and drives or even determines the future of science, technology, the economy, and possibly everything in our world today and tomorrow.

In 2005 in Sydney, we were asked a critical question at a brainstorming meeting about data science and data analytics by several local industry representatives from major analytics software vendors: ``Information science has been there for so long, why do we need data science?'' Related fundamental questions often discussed in the community include ``What is data science?'' \cite{Loukides12}, and ``Is data science old wine in new bottles?'' \cite{Agarwal14}. Data science and relevant topics have become the key concern in panel discussions at conferences in statistics, data mining, and machine learning, and more recently in big data, advanced analytics, and data science. Typical topics such as ``grand challenges in data science'', ``data-driven discovery'', and ``data-driven science'' have frequently been visited and continue to attract wide and increasing attention and debate. These questions are  mainly posted from research and disciplinary development perspectives; while there are many other important questions, such as those relating to data economy and competency, these perspectives are less well considered in the conferences referred to above.

A fundamental trigger for the above questions and many others not mentioned here is the exploration of new or more complex challenges and opportunities \cite{Jagadish14,Cao16ds,Cao10dddm,Khan14} in data science and engineering. Such challenges and opportunities apply to existing fields, including statistics and mathematics, artificial intelligence, and other relevant disciplines and domains that have never been addressed, or have not been adequately addressed, in the classic methodologies, theories, systems, tools, applications and economy of relevant areas. Such challenges and opportunities cannot be effectively accommodated by the existing body of knowledge and capability set without the development of a new discipline. On the other hand, data science is at a very early stage and is engendering enormous hype and even bewilderment; issues and possibilities that are unique to data science and big data analytics are not clear, specific or certain. Different views, observations, and explanations -- some of them controversial -- have thus emerged from a wide range of perspectives.  
        
There is no doubt, nevertheless, that the potential of data science and analytics to enable data-driven theory, economy, and professional development is increasingly being recognized. This involves not only core disciplines such as computing, informatics, and statistics, but also the broad-based fields of business, social science, and health/medical science. Although very few people today would ask the question we were asked 10 years ago, a comprehensive and in-depth understanding of \textit{what data science is}, and \textit{what can be achieved with data science and analytics research, education, and economy} \cite{Cao-uds-2017}, has yet to be commonly agreed.  

Motivated by the above concerns and observations, this article shares the findings from a comprehensive survey of the journey from statistics and data analysis to data science. It constructs an overview of data science as a field in terms of its research, innovation, economy, profession, and education\footnote{Interested readers may refer to a monograph: L. Cao, Understanding Data Science, to be published by Springer soon for comprehensive discussions about data science.}. This is built on (1) our observations and experience in providing real-life data innovation and practices to large government and industry organizations; (2) the education and training opportunities that have been created for professionals at various levels; and (3) reflection on our view on critical issues, future directions and strategic opportunities in data science and analytics. 

Focusing on data science (rather than big data), it is clear that only a few articles and references have discussed its history and contents, such as in Press \cite{Press13}, Donoho \cite{Donoho15}, and Galetto \cite{Galetto16}. A comprehensive review of data science was provided in Donoho \cite{Donoho15}, which focuses on the evolution of data science from statistics. 
To the best of our knowledge, this paper is the first in the field to present such a comprehensive and in-depth survey and overview. Unlike studies which focus on evolution and specific disciplinary perspectives, this paper provides an introduction to the major aspects and domains of data science research, economy, profession, disciplinary development, and education. This overview  complements our other contributions on specific data science issues and perspectives, i.e., the realities and pitfalls in Cao \cite{Cao16np}, the challenges and disciplinary directions in Cao and Fayyad \cite{Cao16ds}, the profession and education of data science in Cao \cite{Cao16pe}, and the book on understanding data science in Cao \cite{Cao-uds-2017}.

There are several key terms, such as data analysis, data analytics, advanced analytics, big data, data science, deep analytics, descriptive analytics, predictive analytics, and prescriptive analytics, which are highly connected and easily confusing. Table \ref{tab:keyterms} lists and explains them, which are also the key terms widely used in this review. A list of data science terminology is available at www.datasciences.org.

\begin{table}%
\tbl{Some key terms in data science.\label{tab:keyterms}}{%
\begin{tabular}{|l|l|}
\hline
Key terms			& Description					 \\ 
\hline
Advanced analytics	&  Refers to theories, technologies, tools and  processes that enable an in-depth\\ 
					&	understanding and discovery of actionable insights in big data, which  \\  
					&	cannot be achieved by traditional data analysis and processing theories, \\
					&	technologies, tools and processes.	\\
\hline
Big data			&  Refers to data that are too large and/or complex to be effectively and/or \\ 							
					&	efficiently handled by traditional data-related theories, technologies and tools. 		\\
\hline
Data analysis		&  Refers to the processing of data by traditional (e.g., classic statistical,\\ 							
					&	mathematical or logical) theories, technologies and tools for obtaining \\									
					&	useful information and for practical purposes. \\ 
\hline
Data analytics		&  Refers to the theories, technologies, tools and processes that enable an \\
					&	in-depth understanding and discovery of actionable insight  into data.  \\
					&	Data analytics consists of descriptive analytics, predictive analytics, \\	 							
					& and prescriptive analytics.	\\
\hline
Data science	&  Is the science of data.	\\
\hline
Data scientist	&  Refers to those people whose roles very much center on data. \\
\hline
Descriptive analytics	&  Refers to the type of data analytics that typically uses statistics to \\
						&	describe the data used to gain information, or for other useful purposes.\\ 					
\hline
Predictive analytics	&  Refers to the type of data analytics that makes predictions about unknown future \\ 														&	events and discloses the reasons behind them, typically by advanced analytics.	\\
\hline
Prescriptive analytics	&  Refers to the type of data analytics that optimizes indications and recommends \\ 														&	actions for smart decision-making.	\\
\hline
Explicit analytics		&  Focuses on descriptive analytics typically by reporting, descriptive\\ 								
						&	analysis, alerting and forecasting.		\\
\hline
Implicit analytics	&  Focuses on deep analytics, typically by predictive modeling, optimization,\\ 							&	prescriptive analytics, and actionable knowledge delivery.		\\
\hline
Deep analytics		&  Refers to data analytics that can acquire an in-depth understanding of why \\
					&	and how things have happened, are happening or will happen, which cannot \\
					&	be addressed by descriptive analytics.		\\
\hline
\end{tabular}}
\end{table}

The paper is organized as follows. Section \ref{sec:evolution} tracks the progression from data analysis to data science, and addresses the fundamental question: \emph{What is data science?} In Section \ref{sec:era}, the main features, initiatives, activities, and status of the era of data science are summarized.  The evolution, state-of-the-art, paradigm shift, 
and major tasks of deep analytics, as the keystone of data science,  are discussed in Section \ref{sec:analytics}. Major challenges and directions of data-driven innovation are presented in Section \ref{sec:innovation}. Section \ref{sec:economy} summarizes new data-driven industrialization and service opportunities. The data science profession, competency, role of data scientists, and course framework are summarized in Section \ref{sec:education}. The future of data science is briefly discussed in Section \ref{sec:dsfuture},
followed by the conclusion of this work.

\section{From Data Analysis to Data Science}
\label{sec:evolution}

This section summarizes the findings of a comprehensive survey, including ours in Cao \cite{Cao16np}, Cao and Fayyad \cite{Cao16ds}, Cao \cite{Cao16pe} and others such as in Press \cite{Press13}, Donoho \cite{Donoho15} and Galetto \cite{Galetto16}), of the journey from data analysis to data science and the evolution of the interest in data science. Subsequently, the question ``What is data science?'' is addressed.

\subsection{The Data Science Journey}
\label{subsec:journey}

It is likely that the first appearance of ``data science'' as a term in literature was in the preface to Naur's book ``Concise Survey of Computer Methods'' \cite{Naur74} in 1974. In that preface,  \emph{data science} was defined as ``the science of dealing with data, once they have been established, while the relation of the data to what they represent is delegated to other fields and sciences.'' Another term, ``datalogy'', had previously been introduced in 1968 as ``the science of data and of data processes'' \cite{Naur68}. These definitions are clearly more specific than those we discuss today. However, they have inspired today's significant move to the comprehensive exploration of scientific content and development.

The evolutionary journey from data analysis \cite{Huber11} to data science started in the statistics and mathematics community in 1962. It was stated that ``data analysis is intrinsically an empirical science'' \cite{Tukey62}\footnote{On this basis, David Donoho argued that data science had existed for 50 years and questioned how/whether data science really differs from statistics \cite{Donoho15}.}. Typical original work on promoting  data processing included \emph{information processing} \cite{Morrell68} and \emph{exploratory data analysis} \cite{Tukey77}. It was suggested that more emphasis needed to be placed on using data to suggest suitable hypotheses to test. This  contributed to the later term of ``data-driven discovery'' in 1989 \cite{KDD89}.  In 2001, an action plan was suggested in Cleveland \cite{Cleveland01} that would  expand the technical areas of statistics toward data science.

Playing a major role in statistics, \emph{descriptive analytics} (also called \emph{descriptive statistics} in the statistics community) \cite{Stewart87} quantitatively summarizes or describes the characteristics and measurements of a data sample or set. Today, descriptive analytics forms the foundation for the default analytical and reporting tasks and tools in typical data analysis and business intelligence projects and systems. 

Our understanding of the roles of data analysis in those early years extended beyond data exploration and processing to the aspiration to ``convert data into information and knowledge'' in 1977 \cite{IASC77}. More than 20 years later, this desire fostered the origin of the currently popular community of the ACM SIGKDD conference, specifically the first workshop on Knowledge Discovery in Databases (KDD for short) with IJCAI'1989 \cite{KDD89}. In KDD and other data mining conferences, ``data-driven discovery'' was adopted as one of key themes of these events. Since then, key terms such as ``data mining'', ``knowledge discovery'' \cite{Fayyad96} and \emph{data analytics} \cite{Vance11} have been increasingly recognized not only in computer science but also in other areas and disciplines. \emph{Data mining and knowledge discovery} is the process of discovering hidden and interesting knowledge from data. Today, in addition to the well-recognized events KDD, ICML, NIPS and JSM, many regional and international conferences and workshops on data analysis and learning have been created. The latest development is the creation of global and regional conferences on data science, especially the IEEE International Conference on Data Science and Advanced Analytics \cite{DSAA}. DSAA has received joint support from IEEE, ACM and the American Statistics Association, in addition to industry sponsorship. The above efforts have ostensibly made data science the fastest growing and most popular computing, statistics and interdisciplinary communities. 

The development of data mining, knowledge discovery, and machine learning, together with the original data analysis and descriptive analytics from the statistical perspective, forms the general concept of ``data analytics''. The initial data analysis focused on processing data. \emph{Data analytics} is the multi-disciplinary science of quantitatively and qualitatively examining data for the purpose of drawing new conclusions or insights (exploratory or predictive), or for extracting and proving  (confirmatory or fact-based) hypotheses about that information for decision making and action. 

Analytics has also become more business-oriented \cite{Kohavi02}. It now extends to a variety of data and domain-specific analytical tasks, such as business analytics, risk analytics, behavior analytics \cite{Cao2012cba}, social analytics, and web analytics (also generally termed ``X-analytics''). Domain-specific analytics fundamentally drives the innovation and application of data science. Both domain-specific and data-specific analytics and theoretical data analytics have together formed the keystone of data science. 

\begin{figure}
\centerline{\includegraphics[width=1.0\textwidth]{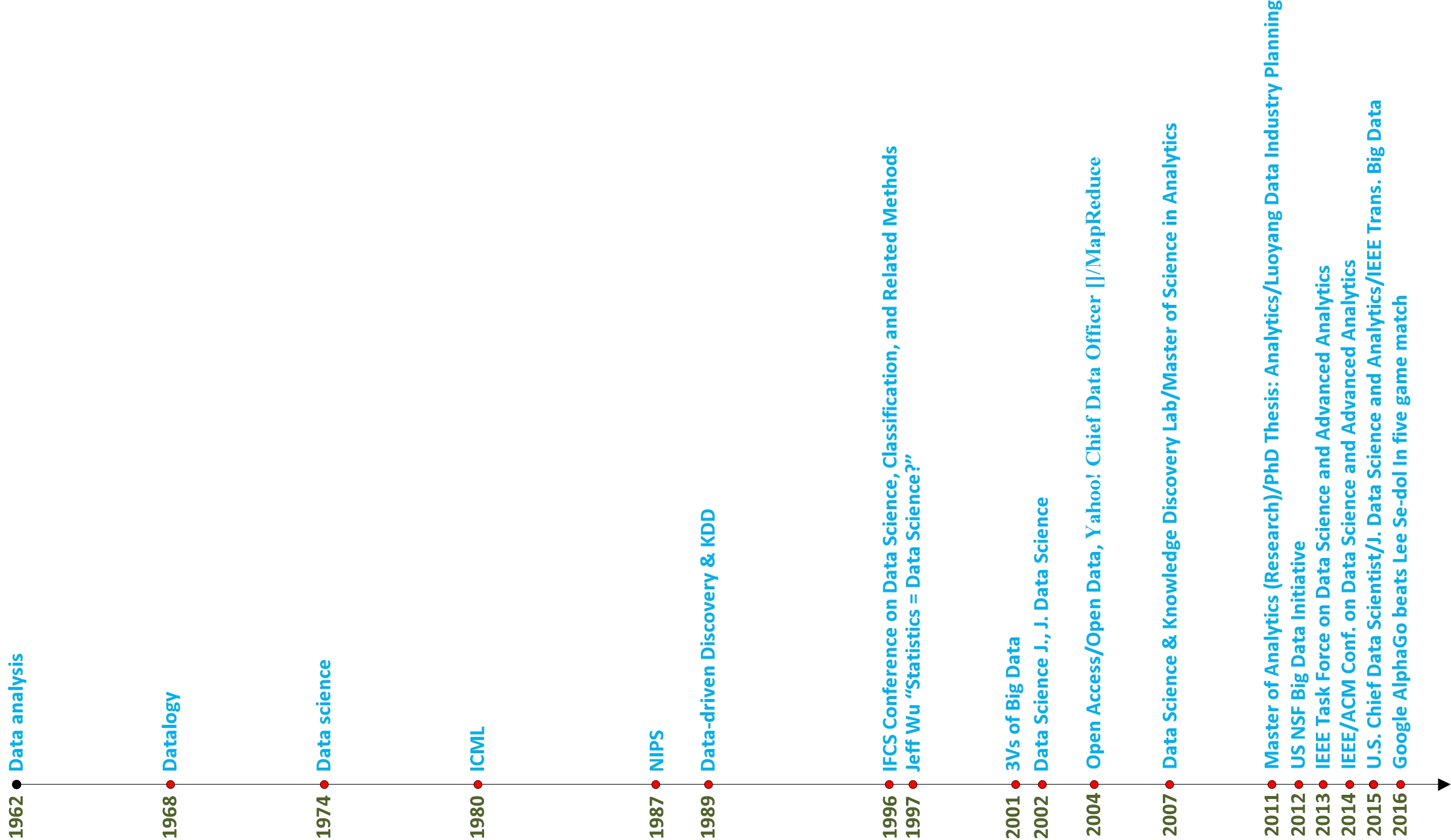}}
\caption{Data science journey (w.r.t. typical events).}
\label{fig:dsjourney}
\end{figure}

Fig. \ref{fig:dsjourney} summarizes the data science journey. It presents the evolution in terms of representative moments, events and major aspects of disciplinary development, government initiatives, scientific agendas, typical socio-economic events, and education.

\subsection{Online Search Interest Trends}
\label{trend}

According to Google Trends \cite{Googletrends}, the online search interest over time in ``data science'' is similar to the interest in ``data analytics'', but is 50\% to 100\% less than the interest in ``big data''. However, the historical search interest in data science and analytics is roughly double the interest shown in big data  about 10 years ago. Compared to the smooth growth of interest in data science and analytics, the interest in big data has experienced a more rapid increase since 2012. When we googled ``data science'', 83.8M records were returned, compared to 365M on ``big data'', and 81.8M on ``data analytics''.

\begin{figure}
\centerline{\includegraphics[width=1.0\textwidth]{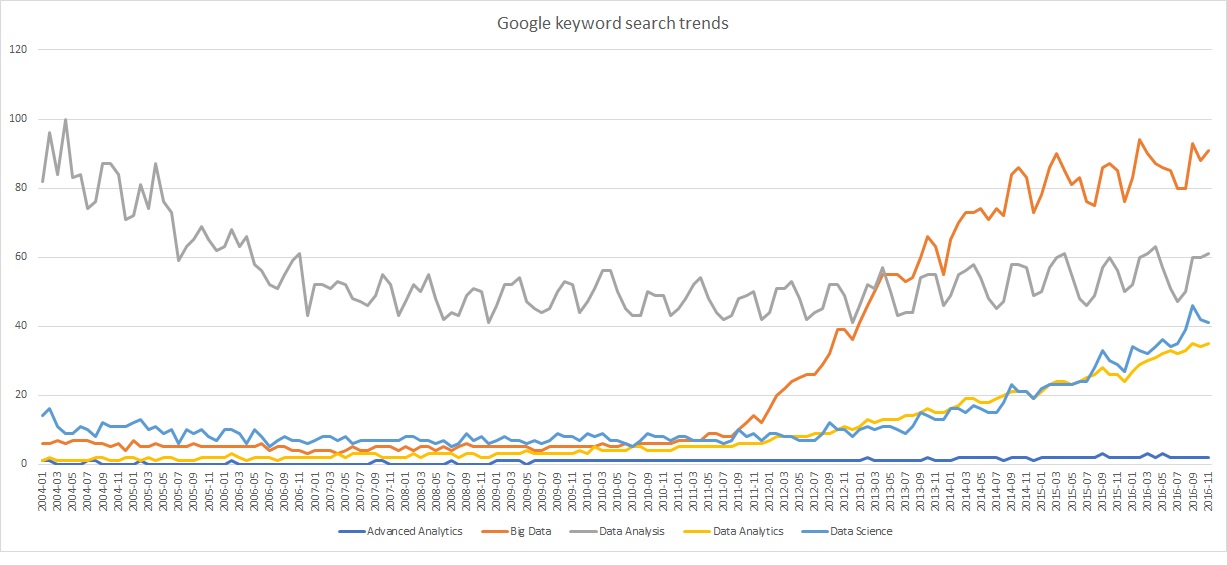}}
\caption{Online search interest trends on data science-related keywords by Google.}
\label{fig:googletrends}
\begin{tabnote}%
\Note{Note:}{The data was collected on 15 November 2016.}
\end{tabnote}%
\end{figure}

Although they do not reflect the full picture, the Google search results in the last 10 years, shown in Fig. \ref{fig:googletrends}, indicate that: (1) Data science, data analysis, and data analytics have much richer histories and stronger disciplinary foundations than big data. (2) The significant boom in big data has been fundamentally business-related, while data science has been highly linked with research and innovation. (3) Data analysis has always been a top concern, although search interest has been flattened and diversified into other hot topics, including big data, data science and data analytics. (4) Interestingly, the word ``advanced analytics'' has received much less attention than all other terms, reflecting the fact that knowledge of, and interest in, more general terms like data analytics is greater than it is for more specific terms such as advanced analytics. (5) Compared to 10 years ago, scrutiny of the search trends in the past four years would find that big data has seen significantly increasing interest from 2012 to 2015 and then less movement; however, the interest in data science and data analytics has consistently increased, although it has grown at a much lower rate (some one third of big data). Data analysis has maintained a relatively stable attraction to searchers during these 10 years.

\subsection{What Is Data Science}
\label{subsec:concept}

The art of data science \cite{Graham12} has attracted increasing interest from a wide range of domains and disciplines. Accordingly, communities or proposers from diverse backgrounds, with contrasting aspirations, have presented very different views or foci. Some examples are that data science is \textit{the new generation of statistics}, is \textit{a consolidation of several interdisciplinary fields}, or is \textit{a new body of knowledge}. Data science also has implications for providing capabilities and practices for the data profession, or for generating business strategies.

Statisticians have had much to say about data science, since it is they who actually created the term ``data science'' and promoted the upgrading of statistics to data science as a broader discipline. This is reflected in a series of earlier actions, such as the following. 
\begin{itemize}
\item Jeff Wu questioned in 1997 whether ``Statistics = Data Science?'' and suggested that statistics should be renamed ``data science'' and statisticians should be known as ``data scientists'' \cite{wu97}. The intention was to shift the focus of statistics from ``data collection, modeling, analysis, problem understanding/resolving, decision making'' to future directions on ``large/complex data, empirical-physical approach, representation and exploitation of knowledge''.
\item William S. Cleveland suggested in 2001 that it would be appropriate to alter the statistics field to data science and ``to enlarge the major areas of technical work of the field of statistics'' by looking to computing and partnering with computer scientists  \cite{Cleveland01}.
\item Leo Breiman suggested in 2001 that it  was necessary to ``move away from exclusive dependence on data models (in statistics) and adopt a more diverse set of tools'' such as algorithmic modeling, which treats the data mechanism as unknown \cite{Breiman01}. 
\item In 2015, a statement about the role of statistics in data science was released by a number of ASA leaders \cite{amstatnews15}, saying that ``statistics and machine learning play a central role in data science.'' Many other relevant discussions are available in AMSTATNEWS \cite{asanews15} and IMS \cite{yu14}.
\end{itemize} 

A large proportion of the conceptual arguments are derived from the data-centric view. For example, data-driven science is mainly interpreted in terms of the reuse of open data \cite{Rust07,OECD04}; data science comprises the numbers of our lives \cite{Miller13}; or data science enables the creation of data products \cite{Loukides12,Loukides11}. In Jagadish \cite{Jagadish15}, six myths were discussed: (1) size is all that matters, (2) the central challenge of big data is that of devising new computing architectures and algorithms, (3) analytics is the central problem of big data, (4) data reuse is low hanging fruit, (5) data science is the same as big data, and (6) big data is all hype. This illustrates the  constituents of the ecosystem, but also shows the divided views within the communities.

Intensive discussions have taken place within the research and academic community about creating data science as an academic discipline \cite{Smith06}. This involves not only statistics, but also a multi-disciplinary body of knowledge that includes computing, communication, management and decision. The concept of data science is correspondingly defined from the perspective of  disciplinary and course development: for example, treating data science as a mixture of statistics, mathematics, computer science, graphic design, data mining, human-computer interaction, and information visualization \cite{Yau09}.

Below, we present several definitions of data science from high-level and disciplinary perspectives, building on the observations and insights we have gained from this review and relevant experience\footnote{Interested readers may refer to Cao \cite{Cao16np} for another definition from the process perspective.}.
\begin{definition}[Data Science$^1$]
A \emph{high-level statement} is: ``data science is the science of data'' or ``data science is the study of data''.
\end{definition}



\begin{definition}[Data Science$^2$]
From the \textit{disciplinary} perspective, data science is a new interdisciplinary field that synthesizes and builds on statistics, informatics, computing, communication, management and sociology to study data and its environments (including domains and other contextual aspects, such as organizational and social aspects) in order to transform data to insights and decisions by following a data-to-knowledge-to-wisdom thinking and methodology.
\end{definition}

Accordingly, a \textit{discipline-based data science formula} is given below: 
\begin{eqnarray}
data~science = statistics + informatics + computing + communication + \nonumber \\
 sociology + management | data + environment + thinking 
\end{eqnarray}
where ``$|$'' means ``conditional on.'' 

The outputs of data science are \textit{data products}  \cite{Loukides11,Loukides12}. We define data products below.
\begin{definition}[Data Products]
A data product is a deliverable from data, or is  enabled or driven by data, and can be a discovery, prediction, service, recommendation, decision-making insight, thinking, model, mode, paradigm, tool or system. The ultimate data products of value are knowledge, intelligence, wisdom and decision.
\end{definition}

The above definition of data product goes beyond technical product-based types and forms in the business and economic domain, such as social network platforms like Facebook, recommender systems like Netflix, and mobile apps like Uber. Data science enables us to explore new data-driven or data-enabled personalized, organizational, educational, ethical, societal, cultural, economic, political, cyber-physical forms, modes, paradigms, innovations, directions and ecosystems, or even thinking, strategies and policies. For example, there is a good possibility that large scale data will enable and enhance the transfer of subjective autonomy to objective autonomy, beneficence and justice in the social sciences \cite{Fairfielda14}. It can enable the discovery of indicators like Google Flu \cite{Lazer14} which may not be readily predicted by domain-driven hypothesis and professionals.

These platforms deliver data products in various forms, ways, channels, and domains that are fundamentally transforming our academic, industrial, governmental, and socio-economic life and world. With the development of data science and engineering theories and technologies, new data products will be created. This creation is likely to take place at a speed and to an extent that greatly exceeds our imagination and thinking, as shown in the evolution of Internet-based products and artificial intelligence systems.  

\section{The Era of Data Science}
\label{sec:era}
In this section, we summarize the main characteristics of data science-related government initiatives, disciplinary development, economy, and profession, as well as activities in these fields, and the progress made to date. Datafication \cite{Ayankoya14}, the quantified self (QS) \cite{Swan13,Clay13,Wolf12,Duncan09,Smarr12,Fawcett16}, initiatives by governments and research institutions 
and open data are discussed as the key drivers of the era of big data and data science.

\subsection{Datafication and Data Quantification}
\label{subsec:datafy}
Data is ubiquitous because  datafication \cite{Ayankoya14} and data quantification are ubiquitous. In addition to the commonly seen data transactions acquired from business and operational information systems, increasingly popular and widespread datafication and data quantification systems and services are significantly strengthening the data deluge and big data realm. Such systems and services include but are not limited to wearables, Internet of Things (IoT), mobile and social applications. 

As we have seen and can predict, datafication and data quantification take place at any time and any place by anybody in any form in any way in a non-traditional manner, extent, depth, variety and speed. 
\begin{itemize}
\item Quantification timing: \textit{anytime quantification}, from working to studying, day-to-day living, relaxing, enjoying entertainment and socializing; 
\item Quantification places: \textit{anyplace quantification}, from biological systems to physical, behavioral, emotional, cognitive, cyber, environmental, cultural, economic, sociological and political systems and environments;    
\item Quantification bodies: \textit{anybody quantification}, from selves to others, connected selves, exo-selves \cite{Kelly12} and the world, and from individuals to groups, organizations and societies;
\item Quantification forms: \textit{anyform quantification}, from observation to drivers, from objective to subjective, from physical to philosophical, from explicit to implicit, and from qualitative to quantitative forms and aspects;
\item Quantification ways: \textit{anysource quantification}, such as sources and tools that include information systems, digitalization, sensors, surveillance and tracking systems, the IoT, mobile devices and applications, social services and network platforms, and wearable \cite{Viseu10} and Quantified Self (QS) devices and services; and 
\item Quantification speed: \textit{anyspeed quantification}, from static to dynamic, from finite to infinite, and from incremental to exponential generation of data objects, sets, warehouses, lakes and clouds.
\end{itemize}

Examples of fast developing quantification areas are the health and medical domains. We are datafying both traditional medical and health care data and ``omics'' data (genomics, proteomics, microbiomics, metabolomics, etc.) and increasingly overwhelming QS-based tracking data \cite{Swan13} on personal, family, group, community, and/or cohort levels.

\subsection{Data Initiatives by Governments}
\label{sebsec:gov}
To effectively understand and utilize everywhere data, data DNA and its potential, increasing numbers of regional and global government initiatives \cite{CIS2015} are being introduced at different levels and on different scales in this age of big data and data science to promote data science research, innovation, funding support, policy making, industrialization, and economy. Table \ref{tab:gov-initiatives} summarizes the major initiatives of several countries and regions.  
\begin{itemize}
\item The Australian Public Service Big Data Strategy \cite{UNpulse} aims to ``provide an opportunity to consider the range of opportunities presented to agencies in relation to the use of big data, and the emerging tools that allow us to better appreciate what it tells us, in the context of the potential concerns that this might raise''. It addresses the identified big data strategy issues \cite{ABDS13}. Australia's whole-of-government Centre of Excellence in Data Analytics \cite{AUbd} coordinates relevant government activities. The Australian Research Council has granted  approval to the Australian Research Council (ARC) Centre of Excellence for Mathematical and Statistical Frontiers \cite{ACEMS} to conduct research on big data-based mathematical and statistical foundations. Another recent effort made by the Australian government was the establishment of Data61 \cite{data61}, which consolidated the relevant data-related human resources in the original National ICT Australia (NICTA) \cite{NICTA} and CSIRO and aims for a unified platform for data research and innovation, engagement with industry and government and academia, and software development.
\item Canada's policy framework Capitalizing on Big Data \cite{CAbd} aims at ``establishing a culture of stewardship ... coordination of stakeholder engagement ... developing capacity and future funding parameters.''
\item China's Guidelines \cite{CNbd} are aimed at boosting the development of big data research and applications, to ``set up an overall coordination mechanism for big data development and application, speed up the establishment of relevant rules, and encourage cooperation between the government, enterprises and institutions.'' China has also set up a national strategic plan for the IoT and big data \cite{CNbd}. Many states and cities in China have launched national big data strategies and action plans for big data and cloud computing \cite{CBDIO,13th5YP}, such as in Beijing \cite{Beijing2016-20}. Probably, a very early example in China was the Luoyang City-sponsored consulting project in 2011, for which we developed a strategic plan for the City's industrial transformation to a ``data industry'' \cite{caoly11}. 
\item The European Union's communication Towards a Thriving Data-driven Economy \cite{EUbd} is ``an action plan to bring about the data-driven economy of the future''. It outlines ``a new strategy on Big Data, supporting and accelerating the transition towards a data-driven economy in Europe. The data-driven economy will stimulate research and innovation on data while leading to more business opportunities and an increased availability of knowledge and capital, in particular for SMEs, across Europe.'' In 2015, the European Data Science Academy, EDSA \cite{EUdsa} was formed.
\item The United Kingdom's Big Data and Energy Efficient Computing initiative funded by the Research Councils UK \cite{UKbd} aims to ``create a foundation where researchers, users and industry can work together to create enhanced opportunities for scientific discovery and development.''
\item The United Nations (UN) Global Pulse Project is ``a flagship innovation initiative of the United Nations Secretary-General on big data. Its vision is a future in which big data is harnessed safely and responsibly as a public good. Its mission is to accelerate the discovery, development and scaled adoption of big data innovation for sustainable development and humanitarian action.'' \cite{UNpulse}
\item The United States (US) Big Data Research Initiative \cite{USNSF} is directed toward ``supporting the fundamental science and underlying infrastructure enabling the big data revolution.'' In 2005, the US National Science Board set the goal that it ``should act to develop and mature the career path for data scientists'' in its report ``Long-lived Digital Data Collections: Enabling Research and Education in the 21st Century'' \cite{NSB05}. In 2009, the Committee on Science of the National Science and Technology Council formed an Interagency Working Group on Digital Data. It published a report  \cite{CSNSTC09} outlining the strategy to ``create a comprehensive framework of transparent, evolvable, extensible policies and management and organizational structures that provide reliable, effective access to the full spectrum of public digital scientific data'', which ``will serve as a driving force for American leadership in science and in a competitive, global information society.'' In addition, the Defence Advanced Research Projects Agency (DARPA) launched its XDATA Program \cite{Xdata}, which aims to develop computational techniques and software tools for processing and analyzing large, imperfect and incomplete data. In 2012, the National Institute of Standards and Technology (NIST) introduced a new data science initiative \cite{Dorr15-2}, and in 2013, the US National Consortium for Data Science was established \cite{USd2d}.
\end{itemize}

\begin{table}%
\tbl{Government initiatives in big data and data science.\label{tab:gov-initiatives}}{%
\begin{tabular}{|l|l|}
\hline
Government	& Representative Initiatives					 \\\hline
Australia	& Public Service Big Data Strategy 	\cite{UNpulse}, Whole-of-Government Centre\\ 
			&   of Excellence on Data Analytics \cite{AUbd}			\\\hline
Canada		& Capitalizing on Big Data 	\cite{CAbd}				\\\hline
China		& Big Data Guideline \cite{CNbd}, China Computer Federation Task Force on Big \\
			&  Data \cite{CCFBDTF}, China National Science Foundation big data program \cite{CNSF}					\\\hline
EU 			& Data-driven Economy \cite{EUbd}, European Commission Horizon 2020	\\
			&  Big Data Private Public Partnership \cite{Horizonbd}				\\\hline
UK			& UK's Big Data and Energy Efficient Computing \cite{UKbd}	\\\hline
UN			& UN Global Pulse Project \cite{UNpulse}						\\\hline
US			& US Big Data Research Initiative \cite{USNSF}, Interagency Working Group \\
			&  on Digital Data  \cite{CSNSTC09}, DARPA's XDATA Program \cite{Xdata}, USA NSF Big \\
			& Data Research Fund \cite{USNSF}		\\\hline
\end{tabular}}
\end{table}

\subsection{The Scientific Agenda of Data Science}
\label{subsec:discipline}
An increasing number of new scientific initiatives, activities and programs have been  created by governments, research institutions, and educational institutions to promote data science as a new field of science.

The original scientific agenda of data science has been driven by both government initiatives and academic recommendations. This was built on the strong promotion of converting statistics to data science, and blending statistics with computing science in the statistics community \cite{wu97,Cleveland01,Iwata08,Hardin15,Hand15,Diggle15,Graham12,Finzer13}. Today, many regional and global initiatives have been taken in data science research, disciplinary development and education, as strategic matters and agenda in the digital era. Several examples are given below. 
\begin{itemize}
\item In Australia, a Go8 report \cite{Brown} suggested the incorporation of data as a keystone in K-12 education through statistics and science by such methods as creating data games for children.
\item In China, the Ministry of Science and Technology very recently announced  the establishment of national key labs in big data research as part of a strategic national agenda \cite{MIST16}. 
\item In the EU, the HLSG report ``Riding the Wave'' \cite{HLSGwave10} and ``The Data Harvest'' \cite{HLSGharest14} urged the European Commission to implement the vision of creating ``scientific e-infrastructure that supports seamless access, use, re-use, and trust of data'' and foster the development of data science university programs and discipline.
\item In the US, a National Science Board report \cite{NSB05} recommended that the National Science Foundation (NSF) ``should evaluate in an integrated way the impact of the full portfolio of programs of outreach to students and citizens of all ages that are `or could be' implemented through digital data collections.'' Different roles and responsibilities were discussed for individuals and institutions, including data authors, users, managers and scientists as well as funding agencies. The report \cite{CSNSTC09} from the US Committee on Science of the National Science and Technology Council suggested the development of necessary knowledge and skill sets by initiating new educational programs and curricula, such as ``some new specializations in data tools, infrastructures, sciences, and management.''
\end{itemize}

An increasing number of research streams, strengths and focused projects have been announced in major countries and regions, including 
\begin{itemize}
\item The US NSF Big Data Research Fund \cite{USNSF}, 
\item The European Commission Horizon 2020 Big Data Private Public Partnership \cite{Horizonbd,EUbd}, and 
\item The China NSF big data special fund \cite{CNSF}.
\end{itemize}

Each of these supports theoretical, basic and applied data science research and development in big data and analytics through respective scientific foundations, high-tech programs and domain-specific funds such as heath and medical funds. Significant investment has been made to create even faster high performance computers. 

Many universities and institutions have either established or are creating research centers or institutes in data science, analytics, big data, cloud computing, and IoT. For example, in Australia, the author created the first data science lab: the Data Science and Knowledge Discovery Lab at UTS in 2007 \cite{DSKD07}, and the first Australian institute: the Advanced Analytics Institute \cite{UTSAAI,ABDS13} in 2011 which implements the RED model of Research, Education and Development (RED) of big data analytics for many major government and business organizations. In the US, top universities have worked on building data science initiatives, such as the Institute for Advanced Analytics at North Carolina State University in 2007 \cite{IAA}, the Stanford Data Science Initiatives in 2014 \cite{Stanford}, and the Data Science Initiatives at University of Michigan in 2015 \cite{MIDS}.

\subsection{Data Science Disciplinary Development}
\label{subsubsec:course}
In contrast to big data that has been driven by data-oriented business and private enterprise, researchers and scientists also  play a driving role in the data science agenda. Migrating from the original push in the statistics communities, various disciplines have been involved in promoting the disciplinary development of data science. This involves the disciplinary structure, intrinsic challenges and directions, course structure and curriculum design, and qualifications for next-generation data scientists \cite{Cao16ds,Cao16np,Cao16pe}. 

In Borne et al. \cite{Borne10}, the authors highlight the need to train the next generation of specialists and non-specialists to derive intelligent understanding from the increased vastness of data from the Universe,  ``with data science as an essential competency'' in astronomy education ``within two contexts: formal education and lifelong learners''. The aim is to manage ``a growing gap between our awareness of that information and our understanding of it.'' In several researches \cite{Fox11,Bailer12,Rudin14,Anderson14-1,Baumer15,Bussaban15}, discussions focus on the needs, variations and addenda of data science-oriented subjects for undergraduate and postgraduate students majoring in mathematics and computing. Case studies of relevant subjects at seven institutions were introduced in Hardin \cite{Hardin15}, with the syllabi collected in Hardin \cite{Hardin47}.

In addition to the promotion activities in core analytics disciplines such as statistics, mathematics, computing and artificial intelligence, the extended recognition and undertaking of domain-specific data science seems to repeat the evolutionary history of the computer and computer-based applications. Data science is warmly embraced by more and more disciplines and domains in which it was traditionally irrelevant, such as law, history and even nursing \cite{Clancy14}. Its core driving forces come from  data-intensive and data-rich areas such as astronomy \cite{Borne10}, neurobiology \cite{Dierick15}, climate change \cite{Faghmous14}, research assessment \cite{Siart15}, media and entertainment \cite{Gold13}, supply chain management (SCM) \cite{Hazena14} and SCM predictive analytics \cite{Schoenherr-jbl13}, advanced hierarchical/multiscale materials \cite{Kalidindi15-2,Gupta15}, and cyberinfrastructure \cite{NSF0728}. The era of data science presents significant interdisciplinary opportunities \cite{Rudin14}, as evidenced by the transformation from traditional statistics and computing-independent research to cross-disciplinary data-driven discovery combining statistics, mathematics, computing, informatics, sociology and management. Data science drives the disciplinary shift of artificial intelligence (AI) from its origins in logics, reasoning and planning-driven machine intelligence to  meta-synthesizing ubiquitous X-intelligence-enabled complex intelligent systems and services  \cite{Qian91,Qian93,Metasynthesis09,Metasynthetic15}. 

A very typical inter-, multi- and cross-disciplinary evolutionary trend is the adoption and adaptation of data-driven discovery and science in classic disciplines from an informatics perspective. This has resulted in the phenomenon  of \emph{X-informatics} for transforming and reforming  the body of knowledge. Typical examples include astroinformatics, behavior informatics \cite{Cao10-1,Cao12-1}, bioinformatics, biostatistics, brain informatics, health informatics, medical informatics, and social informatics, to name a few \cite{Informatics}. Hence, it is not surprising to see courses and subjects being offered in specific areas such as biomedical informatics, healthcare informatics, and even urban informatics.

Following the creation of the world first coursework Master of Science in Analytics \cite{MAS07} created at North Carolina State University in 2007, and the world first  Master of Analytics by Research and PhD in Analytics launched at the University of Technology Sydney in 2011 \cite{UTSA,wired14}, more than 150 universities and institutions have now either created or are planning courses in data science, big data and analytics \cite{Silk}. The majority of these course initiatives focus on training postgraduate specialists and certificate-based trainees in business disciplines, followed by the disciplines of computer science and statistics. 

Several repositories \cite{Silk,DSC,Github,Classcentral,USDSC} collect information about courses and subjects related to analytics, data science, information systems and management, statistics, and decision science.  
For example, according to DSC \cite{DSC} and Github \cite{Github}, there are currently about 500 general or specific subjects or courses  that relate to data analytics, information processing, data mining and machine learning, of which 78\% are offered in the US. Seventy-two percent are offered at Master's level, with only 7\% at bachelor level, and 3.6\% at doctoral level. About 30\% are online courses. 
From the disciplinary perspective, some 43\% of courses  specifically encompass ``Analytics'', compared to 18\% on ``Data Science'' and only 9\% on ``Statistics''. Approximately 40\% focus on business and social science aspects.   
In Classcentral \cite{Classcentral}, 138 courses and subjects are available. A number of US programs are listed in USDSC \cite{USDSC},  most being created in business and management disciplines.  

More than 85\% of courses \cite{DSC} cover a broad scope of big data, analytics, and data science and engineering. Some courses only offer training in very specific technical skills, capabilities and technologies, such as artificial intelligence, data mining, predictive analytics, machine learning, visualization, business intelligence, computational modeling, cloud computing, information quality, and analytics practices. It is very rare to find courses that are dedicated to analytics project management and communication skill training \cite{Faris11}, although several courses  on decision science are offered.

Online data science courses  significantly complement traditional education and typically offer a successful Internet-based data business model.  The corporate training market has seen increasing competition as vendors and universities invest more resources in this area: the SAS training courses are one such example. Online courses such as those offered as a massive open online course (MOOC) and by open universities are quickly feeding the market. 

Today, an increasing number of courses are offered in the MOOC mode \cite{Fox15,Boyer15}, such as through Class Central \cite{Classcentral}, Coursera \cite{Coursera}, edX \cite{Edx}, Udacity \cite{Udacity} and Udemy \cite{Udemy}. The MOOC model is fundamentally changing the way courses are offered by utilizing online, distributed and open data, curriculum development resources and expertise, and delivery channels and services. Course development technologies such as Google Course Builder \cite{Googlecb} and Open edX \cite{OPENedX} are used to create online courses and their operations.  
 
Most of the available courses  focus on classic subjects, in particular statistics, data mining, machine learning, prediction, business intelligence, information management, and database management. New programming languages including R and Python, and cloud infrastructure MapReduce and Hadoop are highlights in these courses. Techniques related to off-the-shelf software and tools are often emphasized. Very few subjects are specified for advanced analytics, real-life analytics practices, communication, project management, and decision-support. An increasing number of courses are created to address domain-specific demands, such as incorporating statistics, business analytics, web and social network analytics into SCM predictive analytics \cite{Schoenherr-jbl13}.

\subsection{New Data Economy and Industry Transformation}
\label{subsec:indtrans}
The recognition of the values and potential  of data science and analytics and its rapid growth have also been driven and promoted by the evolution of a new data economy and industry transformation, such as large private data enterprise. The advancement of data science and big data analytics is conversely significantly influencing and driving the development of a new data economy, industry transformation and increase in productivity. This wave of data economy upgrading and industry transformation features the revolution of advanced artificial intelligence-enabled technologies and businesses, and the complementary advances in AI and the AI-driven data economy are largely propelled by data science and analytics. They include inventing, commercializing, and applying infrastructures, tools, systems, services, applications, and consultations for managing, discovering, and utilizing deep data intelligence and synthesizing X-intelligences and X-complexities \cite{Cao16ds}. 

A typical indicator is the 2010 IBM Global CEO study, from which the resultant report \cite{IBM2010} draws the following conclusions: ``Yet the emergence of advanced technologies like business analytics can help uncover previously hidden correlations and patterns, and provide greater clarity and certainty when making many business decisions.'' To manage the increasing complexity, the CEOs in this study believe that ``a better handle on information and mastery of analytics to predict consequences of decisions could go a long way in reducing uncertainty and in forging answers that are both swift and right.'' This leads to their desire to ``translate data into insight into action that creates business results'' and to ``take advantage of the benefits of analytics.'' 

Today, it can safely be said that data science has enabled the so-called ``new economy'' as evidenced by large private enterprises such as Facebook, Google and Alibaba. This new \emph{data economy} is data product-based and data technology-driven. An increasing number of organizations recognize the value of data as a strategic asset and invest in building infrastructure, resources, talent, and teams to support enterprise innovation, and to create differentiators that will lift competition and productivity. 
Leading Internet-based data-driven businesses \cite{Dhar13}, such as Google, Facebook, SAS, Alibaba, Baidu and Tencent, have overtaken traditional enterprise giants. 

Classic manufacturing-focused and core business-oriented companies, including IBM, Intel \cite{Stonebraker13} and Huawei, have also all launched corresponding initiatives and strategic actions for big data, IoT, and/or cloud computing, and are pursuing the strategy of data product-based transformation, productivity growth and innovation. Data science has been their new innovation engine for productivity and competition upgrade. Core businesses, including banks, capital market firms, telecommunication service providers, and insurance companies, are leading the way in  datafying, quantifying, analyzing and using data. It is encouraging to see that other traditional business sectors, such as agriculture, tourism, retail, property and education, are also investing in data analytics to transform their productivity and competitive advantage. 

Many new start-ups and spin-offs have emerged rapidly in  recent years and have focused on data-based business, products and services. This is reflected in a fast-evolving big data landscape \cite{BDL}, which covers data sources \& API, infrastructure, analytics, cross-infrastructure/analytics, open source initiatives, and applications. Every year, this changing landscape sees significant, swift growth. As a result of datafication and data quantification, new platforms, products, applications, services and economic models such as Spark and Cloudera have quickly emerged in analytics and big data.

\subsection{Data Professional Community Formation}
\label{subsec:community}
The growth and recognition of an emerging field can be effectively measured in terms of  the formation width, depth and speed of its professional communities. The data science and analytics community is growing incredibly quickly.

The first indicator is the emergence of dedicated publication venues in this area. Several journals  on data science have been established. These include the Journal of Data Science \cite{JDS}, launched in 2002, which is devoted to applications of statistical methods at large; the electronic Data Science Journal \cite{DSJ} relaunched by CODATA in 2014; the EPJ Data Science \cite{EPJDS} launched in 2012; the International Journal of Data Science and Analytics (JDSA) \cite{JDSA} in 2015 by Springer; IEEE Transactions on Big Data \cite{TOBD} in 2015; and the Springer Series on Data Science \cite{SSDS} and the Data Analytics Book Series \cite{DABS}. 

Other publications are in development by various regional and domain-specific publishers and groups. Some examples are the International Journal of Data Science \cite{IJDS}, Data Science and Engineering \cite{DSE} published on behalf of the China Computer Federation (CCF) \cite{DSE}, the International Journal of Research on Data Science \cite{IJRDS}, and the Journal of Finance and Data Science \cite{JFDS}. 

The second indicator can be found in the creation of a data science community which is significantly enhanced by conferences, workshops and forums dedicated to the promotion of data science and analytics. There are also many well-established venues which  either focus on specific aspects such as KDD and ICML or have  adjusted their previous non-data and/or analytics focus, such as the  traditional AI conferences IJCAI and AAAI. 
\begin{itemize}
\item The first conference to adopt ``data science'' as a topic was the 1996 IFCS Conference on Data Science, Classification, and Related Methods \cite{IFSC-96}, which  included papers on general data analysis issues. 
\item The IEEE International Conference on Data Science and Advanced Analytics (DSAA) \cite{DSAA} launched in 2014, was probably the first conference series dedicated to both data science and analytics research and practice. Co-sponsored by ACM SIGKDD, IEEE CIS and the American Statistics Association (ASA), it attracted wide and significant interest from statistics, industry, business, IT and professional bodies. The IEEE Conference on Big Data is an event dedicated to broad areas of big data. 
\item Several other domain-specific and regional initiatives have emerged, such as the three initiatives in India, i.e., the Indian Conference on Data Sciences, the International Conference on Big Data Analytics, and the International Conference on Data Science and Engineering. 
\item Several other conference series have been renamed and repositioned from their original focus on topics such as software engineering and service-based computing  to connect with big data and data science, drawing mainly on key topics of interest and participants  from their original areas.
\item Data analytics, machine learning, and big data have eclipsed the original topics of interest in many traditionally non-data and/or analytics conferences, such as IJCAI, AAAI, VLDB, SIGMOD and ICDE. Not surprisingly, some of these venues now frequently  incorporate more than 50\% of papers on data science matters. 
\end{itemize}

The third indicator is the growth and development of professional (online) communities and organizations established publicly or privately to promote big data, analytics and data science research, practices and education, and interdisciplinary communications. For example: 
\begin{itemize}
\item The IEEE Big Data Initiative \cite{IEEEBD} aims to ``provide a framework for collaboration throughout IEEE'', and states that ``Plans are under way to capture all the different perspectives via in depth discussions, and to drive to a set of results which will define the scope and the direction for the initiative.''
\item The IEEE Task Force on Data Science and Advanced Analytics (TF-DSAA) \cite{TFDSAA} was launched in 2013 to promote relevant activities and community building, including the annual IEEE Conference on Data Science and Advanced Analytics. 
\item The International Institute of Data \& Analytics \cite{IDA} aims to bridge the gaps between academia and industry through the  promotion of data and analytics research, education and development. 
\item The China Computer Federation Task Force on Big Data \cite{CCFBDTF} consists of a network of representatives from academia, industry and government, and organizes its annual big data conference with participants from industry and government.
\item Several groups and initiatives promote dedicated activities of analytics and data science. For instance, Datasciences.org \cite{datasciences.org} collects relevant information about data science research, courses, funding opportunities, professional activities, and platforms for collaborations and partnership. The Data Science Community \cite{DSCEU} claims to be the European Knowledge Hub for Bigdata and Datascience. Data Science Central \cite{Dscentral} aims to be the industry's online resource for big data practitioners. The Data Science Association \cite{Datascienceassn} aims to be a ``professional group offering education, professional certification, conferences and meetups'' \cite{Galetto16}, and even offers a ``Data Science Code of Professional Conduct.'' 
\item Many existing consulting and servicing organizations have adjusted their scope to cover analytics, where they previously focused on other disciplinary matters. Interdisciplinary efforts have been made to promote cross-domain and cross-disciplinary activities and growth opportunities. Examples include INFORMS \cite{INFORMS}, Gartner, McKinsey, Deloitte, PricewaterhouseCoopers, KPMG, and Bloomberg.
\end{itemize}

Lastly, multinational vendors, online and new economy giants, and service providers play a critical driving role in community outreach. Each of these has launched relevant initiatives, such as those by SAS \cite{SAS16}, IBM \cite{IBMbd}, Google \cite{Google} and Facebook \cite{FBdata}. Many professional  interest groups have been set up in social media, including Google groups, LinkedIn, Facebook and Twitter, and are among the most attractive and popular venues  for big data, data science and analytics professionals to share and network.

\subsection{The Open Model and Open Data}
\label{subsec:opendata}

A key feature differentiating the data science era from the previous era lies in the overwhelming adoption and acceptance of the open model rather than a closed one. The \textit{open model} enables free, distributed, and collaborative modes in every aspect of economy, society, research, and living. It supports the innovation of social media like Facebook and LinkedIn, the migration of mobile to smart phone-embedded applications, and industrial transformation such as the migration of physical shop centre-based commerce to online businesses like Taobao.

Typically, open data and data sharing programs have been announced in many countries and domains, such as the US Government open data site \cite{usod}, the UK open data project \cite{UKod,UKpd}, the Australian Government open government data site \cite{AUopengov10,AUopengov13} and  Data Matching program \cite{AUdm90}, and the European Union Open Data Portal \cite{EUod} and data sharing projects \cite{HLSGharest14}. In addition, many Open Access schemes are increasingly being accepted by academic journals. 

Efforts have also been made in diverse societies to create shareable data repositories, especially for science and research. Examples of open repositories are the global climate data \cite{Climatedata}, the global terrorism database \cite{Terrorismdata}, the Yahoo Finance data \cite{Yahoofinance}, the Gene Expression Omnibus \cite{GEOdata}, mobile data \cite{Mobiledata}, the UCI repositories for machine learning \cite{UCIdata}, the Linguistic Data Consortium data for Natural Language Processing \cite{LDC15}, the TREC data for text retrieval \cite{Trec15}, Kaggle competition data \cite{Kaggledata}, and the VAST challenge \cite{Vast15} for visual analytics, to name a few.

\section{Data Analytics: A Keystone of Data Science}
\label{sec:analytics}
In the age of analytics, what is to be analyzed, 
what constitutes the analytics spectrum for understanding data, and what form the paradigm shift of analytics takes are critical questions to be answered. 
We  address these issues in this section.

Data and analytics form a comprehensive map that covers 
\begin{itemize}
\item the whole life cycle of the data from the past to the present and the future, 
\item the analytics from explicit (known) analytics and reactive understanding  to implicit (unknown) analytics and proactive early prediction and intervention, and 
\item the journey from data exploration (by descriptive and predictive analytics) to the delivery of actionable insights and decisions through prescriptive analytics and actionable knowledge delivery \cite{dddm10}. 
\end{itemize}

\subsection{Data-to-Insight-to-Decision Whole-of-Life Analytics}
\label{subsec:lifecycle}

\begin{figure}
\centerline{\includegraphics[width=0.95\textwidth]{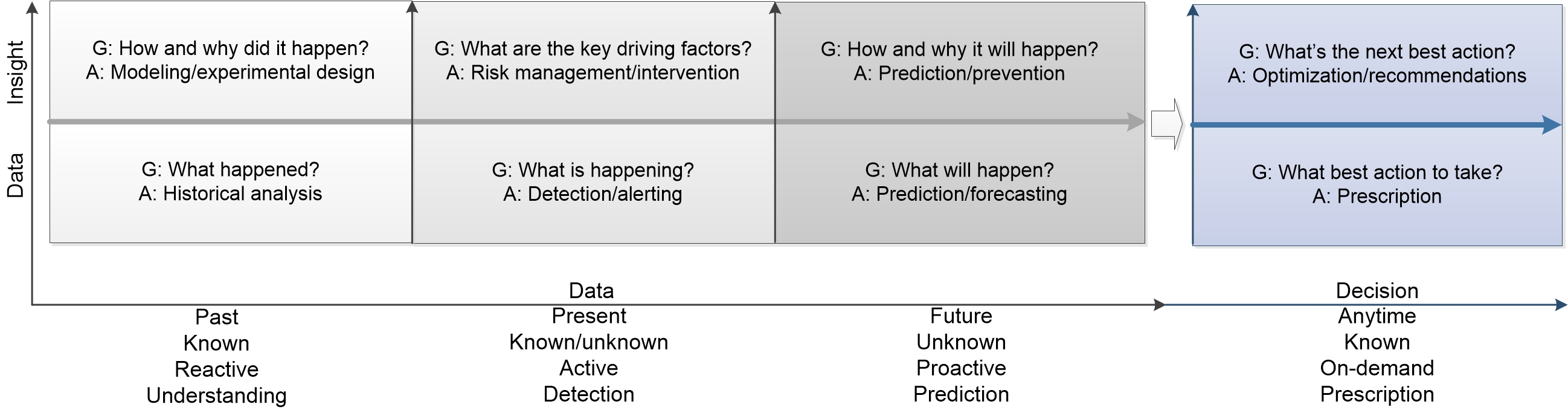}}
\caption{Data-to-insight-to-decision whole-of-life analytics.}
\label{fig:one}
\end{figure}

As shown in Fig. \ref{fig:one}, the data-to-insight-to-decision transfer at different time periods and analytic stages is embodied along the whole-of-life analytics. This can be further represented in terms of a variety of analytics goals (G) and approaches (A) to achieve the data-to-decision goal. 
\begin{itemize}
\item	\emph{Past data}: the main focus of historical analytics is to explore ``what happened'' in the data and business, and to gain insights into ``how and why it happened'' through modeling and experimental design, etc. This stage focuses on ``we know what we know'' to conduct a reactive understanding of what took place. 
\item	\emph{Present data}: detection  at this stage is mainly focused on exploring ``what is happening'', to generate insights about ``how and why it happens''. This stage addresses ``we know what we do not know'' with alerts generated about suspicious events, or interesting groups or patterns presented in the data and business. The insights are extracted for decision-making purposes, such as real-time risk management and intervention, to address the question ``what are the key driving factors?''
\item	\emph{Future data}: predictive analytics is undertaken to investigate ``what will happen'' in the future, and to achieve insights into ``how and why it will happen'' by estimating the occurrence of future events, grouping and patterns. The aim of this stage is to solve the problem that ``we do not know what we do not know'' by achieving proactive understanding, forecasting and prediction, and early prevention. 
\item	\emph{Actionable decision}: prescriptive analytics and actionable knowledge delivery are undertaken to investigate ``what best action to take'' to interpret findings from the past, present or future data. This achieves insights into ``what is the next best action'' and enables the corresponding optimal actions and recommendations to be undertaken based on the findings. The aim of this stage is to solve the problem of ``how to actively and optimally manage the problems identified'' by making optimal recommendations and actionable interventions.
\end{itemize}

\subsection{Explicit-to-Implicit Analytics Evolution}
\label{subsec:spectrum}
\begin{figure}
\centerline{\includegraphics[width=0.95\textwidth]{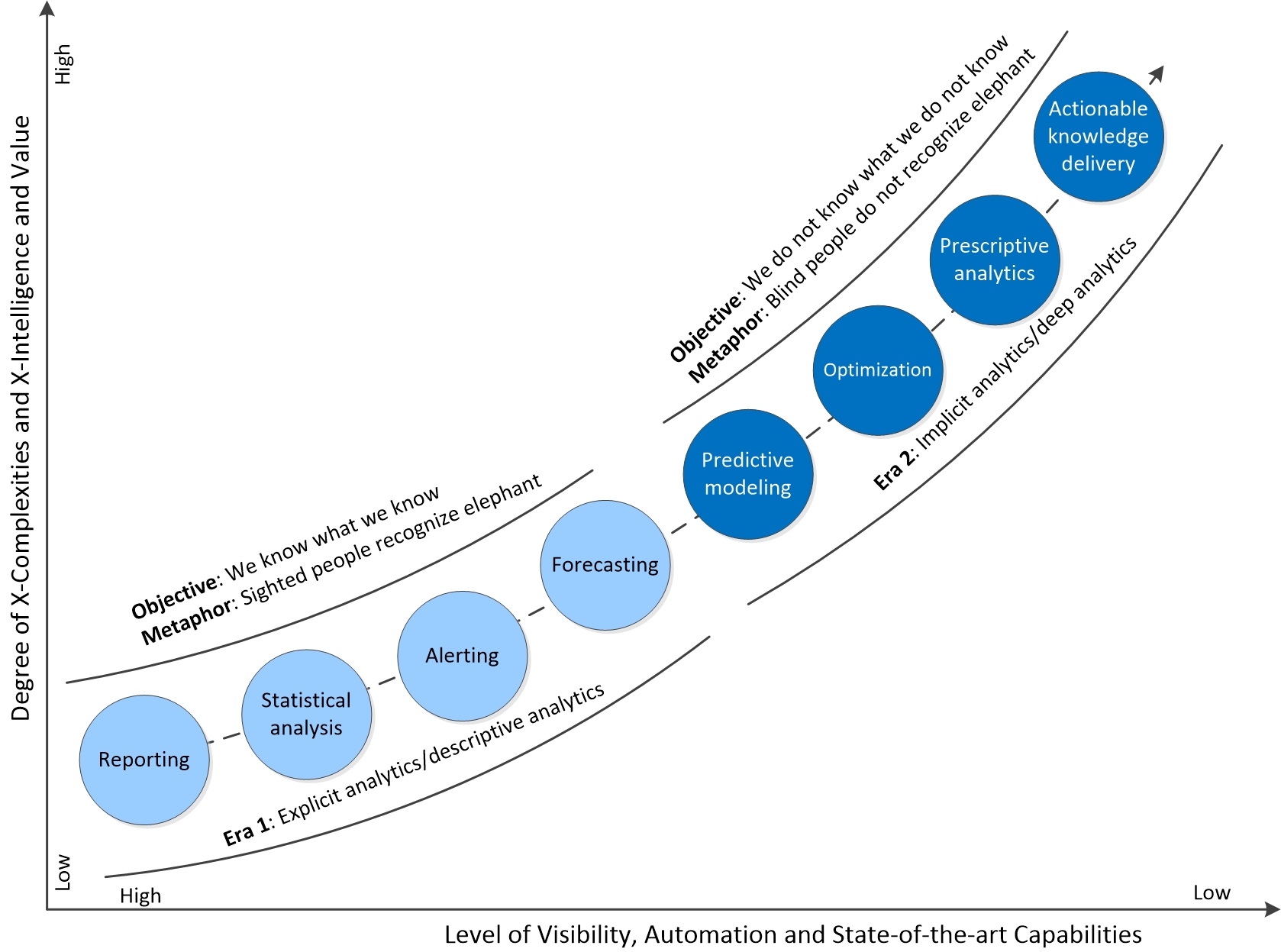}}
\caption{Explicit-to-implicit analytics spectrum and evolution.}
\label{fig:two}
\end{figure}

As discussed in Section \ref{sec:evolution}, the last four decades  have seen the transfer of data analysis on small and simple data, along with hypothesis testing, to data analytics on large and complex data for hypothesis-free knowledge and insight discovery. Today, the significance and innovation of analytics are better recognized than at any previous time. Correspondingly, a critical question to ask is \emph{What is the conceptual map and evolution of data analytics?} Fig. \ref{fig:two} shows a high-level conceptual view of the spectrum and evolution of analytical components and tasks in terms of two major dimensions. 
\begin{itemize}
\item Dimension 1: Levels of visibility, automation and state-of-the-art capabilities: i.e., the level of data and analytics complexity that is visible to users, the level of automated data analytics, and the level of available capability to handle the complexity and support the automation. With the upgrading of analytics, the visibility of data and analytics  becomes lower and the level of automated data analytics is lower too. As data complexity increases, the available capability is weakened. The goal of analytics is to increase the visibility, automation and capability levels of data understanding, production and application.  
\item Dimension 2: Degree of X-complexities, X-intelligence and value: i.e., the degree of data complexity and X-intelligence involved in data and analytics are increased with the movement from lower-level analytics to higher-level analytics. During this process, the level of learned intelligence and value resulting from the corresponding analytics is increased. 
\end{itemize}

As shown in Fig. \ref{fig:two}, there are many typical analytical approaches and components that may be involved in executing analytics tasks. They include reporting, statistical analysis, alerting, forecasting, predictive modeling, optimization, prescriptive analytics, and actionable knowledge delivery (delivering insights-based actions for business decision-making and operations) \cite{dddm10}. The listed approaches may be used for non-analytical purposes, and the corresponding analytical tasks may be addressed by non-analytical approaches. An example is optimization, which may be used for analytics to select the best options as an analytics approach or may be achieved by findings from analytics approaches as an analytics objective. There may be different foci and connections between the listed analytics approaches. For example, forecasting may be used as an approach for prediction when it focuses on probabilistic estimates of possible futures, while prediction may involve broad techniques and objectives for estimating outcomes. 

We also roughly categorize the multiple components and tasks in analytics evolution into two main eras from the perspective of the disciplinary development of analytics:
\begin{itemize}
\item Era 1: The era of explicit analytics: which focuses on descriptive analytics. Typical analytics approaches consist of reporting, statistical analysis, alerting and forecasting. 
\item Era 2: The era of implicit analytics: which focuses on deep analytics. Typical analytics approaches are predictive modeling, optimization, prescriptive analytics, and actionable knowledge delivery.
\end{itemize}

Note that Fig. \ref{fig:two} only shows an evolution path of the analytics family. It does not indicate the path of analytics within a specific organization that conducts and utilizes analytics. It does not indicate a linear path of analytics evolution either. Often, a back-and-forth iterative approach is taken in an analytics team, and multiple analytics components may be involved in parallel for exploring multifaceted observations and understandings. 

\subsubsection{The Era of Explicit Analytics: Descriptive Analytics}
\label{subsubsec:explicit}
Typical elements and tasks have in the past focused on explicit descriptive analysis and have the following features:
\begin{itemize}
\item	\emph{Goal}: \emph{we know what we know}, and therefore aim to identify and describe the distribution, generation and trends of data and business problems;  
\item	\emph{Nature of problem}: similar to \emph{sighted people recognize an elephant}, we know what is to be analyzed by hypothesis-based approaches, and for what purposes;
\item	\emph{Approach}: domain-driven analysis for which hypotheses are available from domain-specific knowledge and experts; data analysis tests such hypotheses, and the data verifies and explains the hypotheses;
\item	\emph{Outcome}: focused methods are available from mathematics and statistics as well as from computing. Such methods describe and present what has happened, is happening or will happen in usually small or highly manipulated data.
\end{itemize}

\subsubsection{The Era of Implicit Analytics: Deep Analytics}
\label{subsubsec:implicit}
The limitation of explicit analytics has recently been more widely recognized in the analytics community, such as in handling latent, uncertain and non-IID data \cite{noniid14,Cao15cl}. As a result, the focus has recently shifted to \emph{implicit analytics}, and towards \emph{deep analytics}. Deep analytics gains an in-depth understanding of why and how things have happened, are happening or will happen. Such whys and hows cannot be addressed by descriptive analytics and can determine the next best or worst situation, as well as devise optimal intervention strategies.
\begin{itemize}
\item	\emph{Goal}: \emph{we do not know what we do not know}, and therefore aim to gain a latent but genuine understanding of data and business problems from visible and invisible sources;
\item	\emph{Nature of problem}: similar to \emph{blind people recognize an elephant}, we do not know what is to be analyzed, or even why and what we can obtain;
\item	\emph{Approach}: data-driven discovery by which interesting but hidden insights are learned from data; data creates a view invisible to us and explains the unseen reasons or indicators, to complement domain-driven hypotheses and observations;
\item	\emph{Outcome}: the focus is  on gaining an in-depth, intrinsic and complete understanding of invisible insights, knowledge and wisdom from data, behaviors and environment about what has happened, is happening or will happen in data and business.
\end{itemize}  

Table \ref{tab:one} summarizes the key categories and features of explicit analytics vs. implicit analytics.

\begin{table}%
\tbl{Explicit-to-implicit analytics.\label{tab:one}}{%
\begin{tabular}{|l|l|l|}
\hline
Categories	& Explicit analytics					& 	Implicit analytics \\\hline
Nature		& Sighted people recognize an elephant	&	Blind people do not recognize an elephant\\\hline
Goal		& We know what we know					&	We do not know what we do not know\\\hline
Approach	& Hypothesis + data						&	Data + environment (incl. domain)\\\hline
Outcome 	& Description of data					&	In-depth representation of data\\\hline
\end{tabular}}
\end{table}%

\subsection{Descriptive-to-predictive-to-prescriptive Analytics Paradigm Shift}
\label{subsec:shift}
The paradigm shift from data analysis to data science constitutes the so-called ``new paradigm'' \cite{Nelson09,4thparadigm}, i.e., data-driven discovery. The  history of  analytics from the spectrum and dynamics perspective spans two main eras of analytics, as shown in Fig. \ref{fig:one}. Analytics practices have seen a significant paradigm shift across three major stages: (1) Stage 1: descriptive analytics and reporting, (2) Stage 2: predictive analytics and business analytics, and (3) Stage 3: prescriptive analytics and decision making.

We briefly discuss these three stages below.
\begin{itemize}
\item \emph{Stage 1: Descriptive analytics and business reporting}: the major effort is on explicit analytics, which focuses on descriptive analytics and regular and ad hoc reporting. Limited effort is made on implicit analytics for hidden knowledge discovery, which is mainly achieved by using off-the-shelf tools and built-in algorithms. Business reports (often analytical reports) generated by dashboards and automated processes are the means for carrying findings from analytics to management. 
\item \emph{Stage 2: Predictive analytics and business analytics}: the major effort is on implicit analytics, which focuses on predictive modeling and business analytics (here \textit{business analytics} refers to an in-depth understanding of business through deep analytics. Note that this meaning differs from the broad meaning widely adopted in business and management), with more effort being made to apply forecasting, data mining and machine learning tools for business understanding and prediction. Patterns, scoring and findings are presented through dashboards and analytical reports to management.    
\item \emph{Stage 3: Prescriptive analytics and decision making}: the major effort is on the delivery of recommended optimal (next best) actions for business decisions by discovering invisible knowledge and actionable insights from complex data, behavior and environment. This is achieved by developing innovative and effective customized algorithms and tools to deeply and genuinely understand domain-specific data and business. Consequently, prescriptive decision-taking strategies, business rules, actions and recommendations are disseminated to decision-makers for the purpose of taking corresponding actions. By contrast, relatively limited effort is made on explicit analytics since they are conducted through automated processes and systems.
\end{itemize}

\begin{figure}
\centerline{\includegraphics[width=0.85\textwidth]{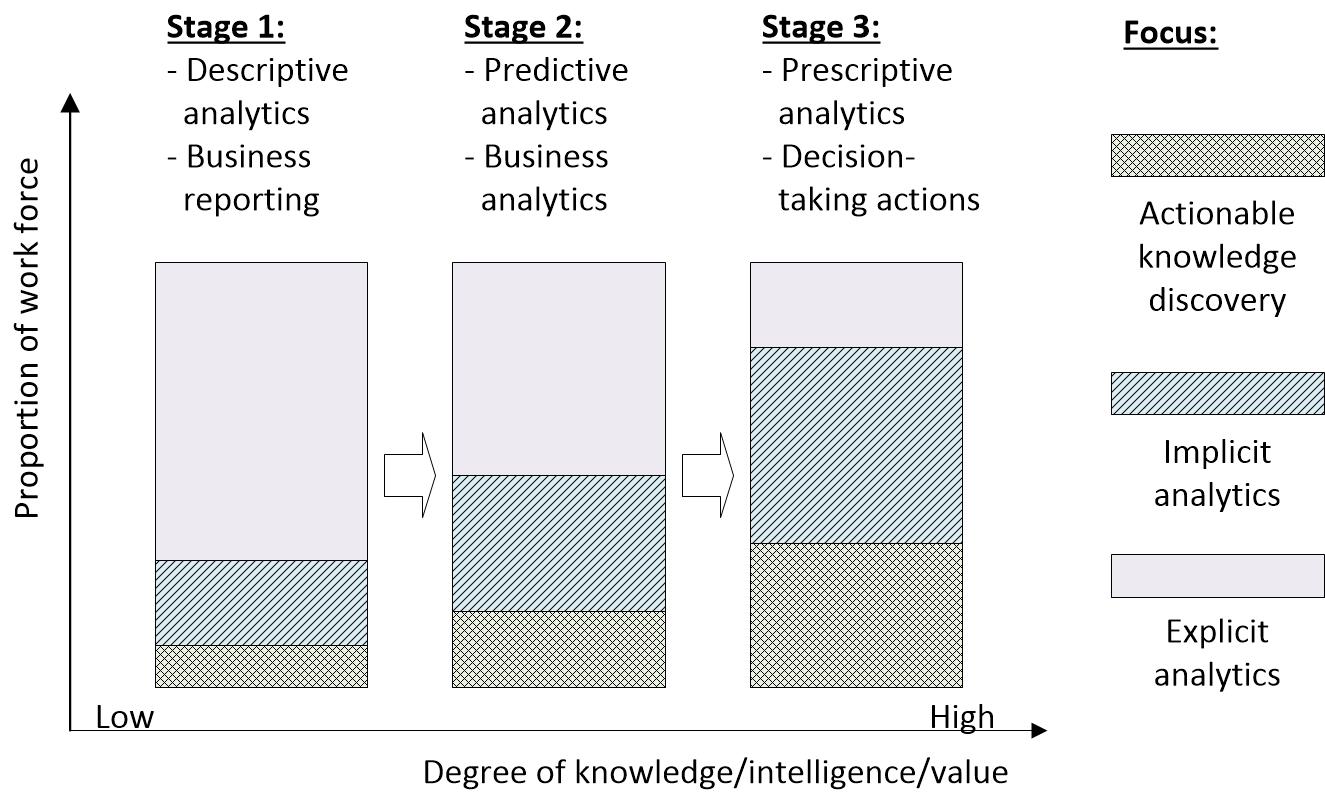}}
\caption{Descriptive-to-predictive-to-prescriptive analytics paradigm shift.}
\label{fig:three}
\end{figure}

During the paradigm shift (as shown in Fig. \ref{fig:three}), a significant decrease is seen in the effort made in routine explicit analytics, which is increasingly undertaken by  automated analytics services. By contrast, a significant increase in effort is seen in implicit analytics and actionable knowledge delivery \cite{dddm10}. The shift from a lower stage to a higher stage accommodates an increasingly higher degree of knowledge, intelligence and value to an organization.  

\section{Data Innovation: Challenges and Opportunities}
\label{sec:innovation}
In this section, we summarize the major challenges and opportunities relating to data science in the relevant communities.

There are two ways of exploring major research challenges: one is to summarize the concerns of the relevant communities, and the other is to scrutinize the issues from the perspective of the intrinsic complexity and nature of data science problems as complex systems \cite{ocis08,Metasynthetic15}. The first approach summarizes the main topics and issues identified in the statistics \cite{Chambers93,wu97,amstatnews15}, informatics and computing \cite{Rudin14,Cao16ds} communities, vendors \cite{Stonebraker13} government initiatives \cite{USNSF,UNpulse,CNbd,ECbd14,UKbd} and research institutions \cite{UTSAAI,IAA}  which focus on data science and analytics. This can result in a picture of the main research challenges. The second approach is much more challenging. It requires us to explore the nature of complex data science problems, and the unknown space of the complexities and comprehensive intelligence in complex data systems. 

\begin{figure}
\centerline{\includegraphics[width=0.85\textwidth]{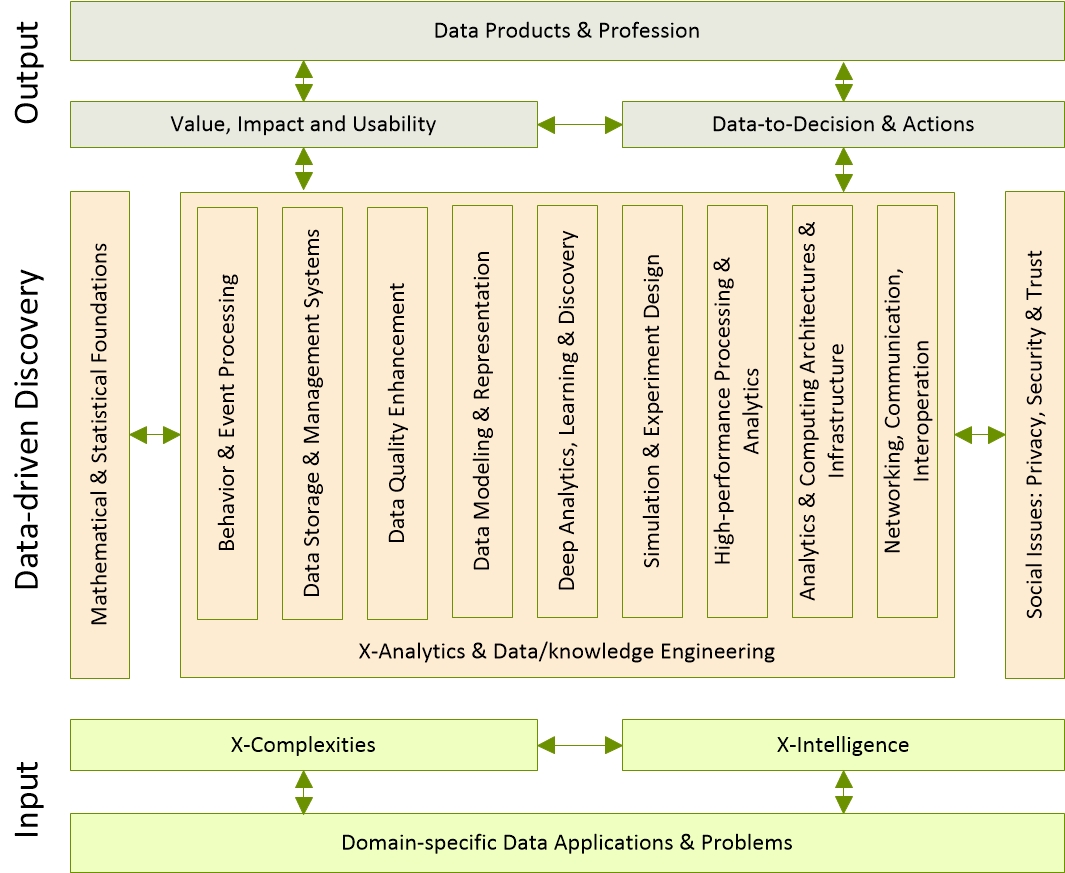}}
\caption{Data science conceptual map.}
\label{fig:seven}
\end{figure}

Fig. \ref{fig:seven} presents a comprehensive conceptual map of data science as a complex system. It summarizes some of the main challenges faced by the data science community in addressing big data complexities \cite{Cao16ds}.
We categorize the challenges facing domain-specific data applications and problems in terms of five major areas: 
\begin{itemize}
\item \textit{Challenges in data/business understanding}: The challenges here are to identify, specify, represent and quantify comprehensive complexities, known as X-complexities \cite{Cao16ds,Metasynthetic15}) and intelligence, known as X-intelligence \cite{Cao16ds,Metasynthetic15}). Such X-complexities and X-intelligence cannot be managed well by existing theories and techniques. However, they nonetheless exist and are embedded in domain-specific data and business problems. The issue is to understand in what form, at what level, and to what extent they exist, and to understand how the respective complexities and intelligence interact and integrate with one another. An in-depth understanding of X-complexities and X-intelligence would subsequently result in devising effective methodologies and technologies for incorporating them into data science tasks and processes.  
\item \textit{Challenges in mathematical and statistical foundations}: The challenges here are to discover and explore whether, how and why existing theoretical foundations are insufficient, missing, and problematic in disclosing, describing, representing, and capturing the above complexities and intelligence and obtaining actionable insights. 
\item \textit{Challenges in X-analytics and data/knowledge engineering}: The challenge is to develop domain-specific analytic theories, tools and systems that are not yet available in the body of knowledge. They will represent, discover, implement and manage the relevant and resultant data, knowledge and intelligence, and support the engineering of big data storage and management, behavior and event processing. 
\item \textit{Challenges in social issues}: This challenge is to identify, specify and respect social issues related to the domain-specific data and business understanding and data science processes, including processing and protecting privacy, security and trust and enabling social issues-based data science tasks, which have not so far been handled well.
\item \textit{Challenges in data value, impact and usability}: This challenge is to identify, specify, quantify and evaluate the value, impact, utility and usability of domain-specific data that cannot be addressed by existing theories and systems, from technical, business, subjective, and objective perspectives.
\item \textit{Challenges in data-to-decision and actions}: The challenge recognized here is the need to develop decision-support theories and systems that will enable data-driven decision generation, insight-to-decision transformation, as well as decision-making action generation, and data-driven decision management and governance. These cannot be managed by existing technologies. 
\end{itemize}

The challenges in X-analytics and data/knowledge engineering involve many specific research issues that have not been properly addressed, for example:
\begin{itemize}
\item \textit{Behavior and event processing}: how to capture, store, model, match, query, visualize and manage behaviors and events and their properties, behavior sequences/streams, and the impact and evolution of behaviors and events of individuals and groups in the physical world. 
\item \textit{Data storage and management systems}: how to design effective and efficient storage and management systems that can handle big data with high volume, velocity and variety, and support real-time, online, and on-the-fly processing and analytics; and how to house such data in an Internet-based (including cloud) environment.
\item \textit{Data quality enhancement}: how to handle both existing data quality issues, such as noise, uncertainty, missing values and imbalance which may be present at very different levels due to the significantly increased scale, extent and complexity of data. At the same time, how to handle new data quality issues emerging in the big data and Internet-based data/business environment, such as cross-organizational, cross-media, cross-cultural, and cross-economic mechanism data science problems. 
\item \textit{Data modeling, learning and mining}: how to model, learn, analyze and mine data that is embedded with comprehensive complexity and intelligence.
\item \textit{Deep analytics, learning and discovery}: how to discover unknown knowledge and intelligence hidden in the space D in Fig. 1 (unknown complexities, knowledge and intelligence, see Can and Fayyad  \cite{Cao16ds}) through inventing new theories and algorithms for implicit and deep analytics that cannot be handled by existing latent learning and descriptive and predictive analytics. Also, how to integrate data-driven and model-based problem-solving which balances common learning models/frameworks and domain-specific data complexity and intelligence-driven evidence learning.
\item \textit{Simulation and experimental design}: how to simulate the complexity and intelligence, working mechanisms, processes, dynamics and evolution in data and business, and how to design experiments and explore the subsequent impact if certain data-driven decisions and actions are undertaken in a business.
\item \textit{High-performance processing and analytics}: how to support large scale, real-time, online, high frequency, Internet-based (including cloud-based) cross-organizational data processing and analytics while balancing local and global resource involvement and objectives. This requires new batch, array, memory, disk storage and processing technologies and systems, and massive parallel processing and distributed/parallel and high-performance processing infrastructure, as well as cloud-based processing and storage. It also requires large and complex matrix calculation, mixed data structures and management systems, and data-to-knowledge management. 
\item \textit{Analytics and computing architectures and infrastructure}: how to facilitate the above tasks and processes by inventing efficient analytics and computing architectures and infrastructure based on memory, disk, cloud and Internet-based resources and facilities.
\item \textit{Networking, communication and interoperation}: how to support the networking, communication and inter-operation between different data science roles in a distributed data science team and during the whole-of-cycle of data science problem-solving. This requires the distributed cooperative management of projects, data, goals, tasks, models, outcomes, workflows, task scheduling, version control, reporting and governance. 
\end{itemize}

Systematic and interdisciplinary approaches and methodologies are required to address the above issues in data science and analytics. These may involve developing a synergy of several research disciplines and areas, including data representation, preprocessing and preparation,  information processing, parallel processing, distributed systems, high performance computing, data management, data warehousing, cloud computing, evolutionary computation, neural networks, fuzzy systems, enterprise infrastructure, system architecture, communication and networking, integration and interoperation, machine learning, data modeling, analytics and mining, service computing, system simulation, experimental design, and evaluation. It may also involve business and social aspects, including industry transformation, enterprise information systems, business intelligence, business process management, project management, information security, trust and reputation, privacy processing, business impact modeling, business value,  and utility evaluation \cite{DSAA,cao16-1}. Inter-disciplinary initiatives are necessary to bridge the gaps between the respective disciplines, and to create new opportunities for the invention and development of new technologies,  theories and tools to address critical complexities in complex data science problems that cannot be addressed by singular disciplinary efforts.

\section{Data Economy: Data Industrialization and Services}
\label{sec:economy}
Data science and big data analytics have led to next-generation economy innovation, competition and productivity \cite{Mckinsey11}, as shown by the rapidly updated Big Data Landscape \cite{bdl16}. Significant new business opportunities and previously impossible prospects have become possible through the creation of data products, data economy and data industrialization and services \cite{yiu12,CSCbd,IBMbd,Loukides12}. In this section, we discuss such opportunities. 

\subsection{Data Industry}
\label{subsec:industry}
If  data is viewed in the same way as  oil, as the new international currency, then clearly the global economy is experiencing a revolutionary change from data poor to data rich and data-driven. On one hand, data industrialization creates new business, where companies, organizations and even countries compete over how to best use data to create new data products. On the other hand, core businesses, including retail business and manufacturing, are giving way to a new economy that is centered on the data industry and digital economy. This is evidenced by the domination of data-enabled companies listed in the top 10 global companies, especially the largest data company, Google, and the largest Initial Public Offering Alibaba. 

\begin{figure}
\centerline{\includegraphics[width=0.95\textwidth]{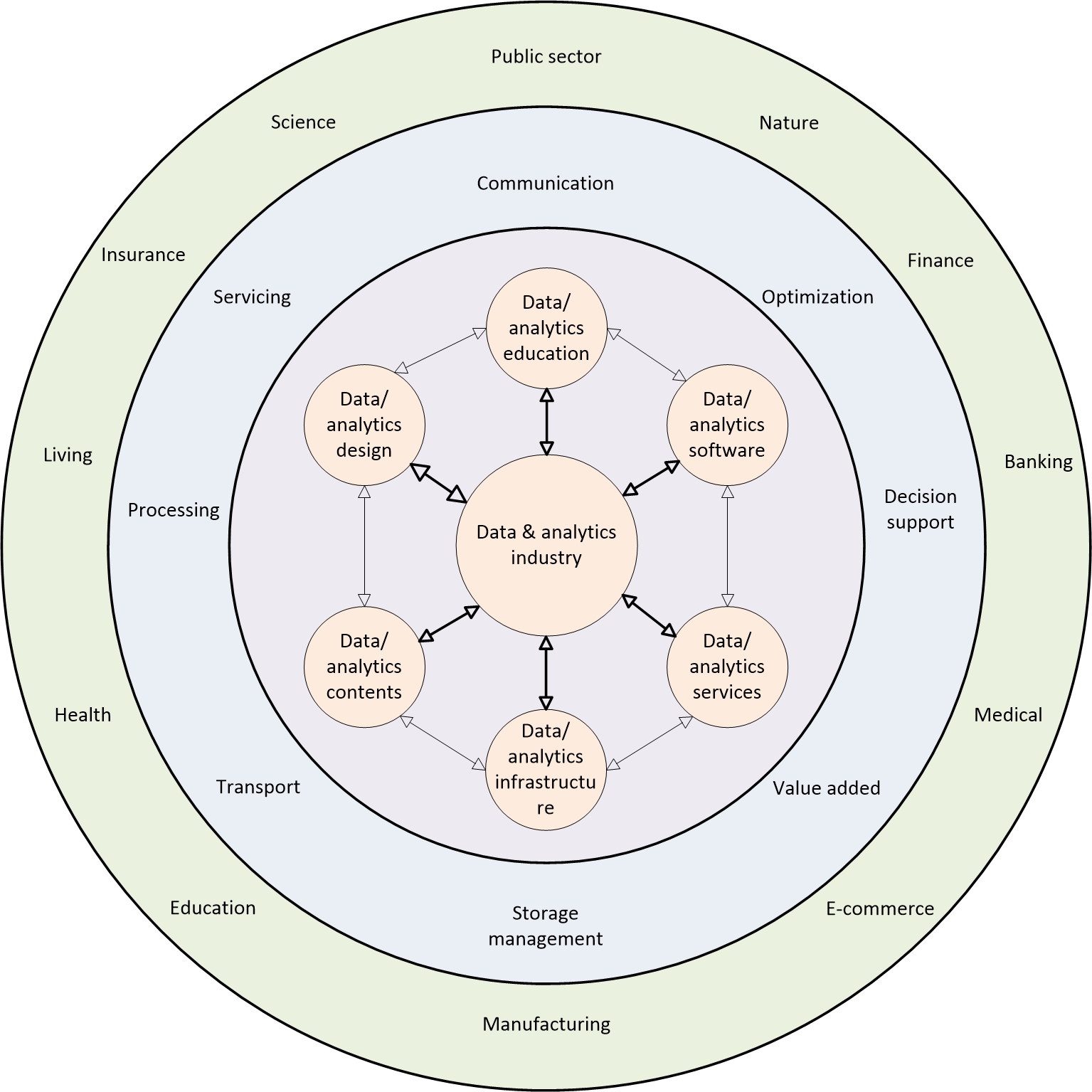}}
\caption{Data \& analytics-enabled industry and business transformation.}
\label{fig:ten}
\end{figure}

The data industry is taking  shape and gaining significance as a driving force in the new global economy. Without loss of generality, Fig. \ref{fig:ten} illustrates those aspects in which new data business and  the resultant areas of data business may grow. The main driving forces of the data industry come from the following six core areas: data/analytics design, data/analytics content, data/analytics software, data/analytics infrastructure, data/analytics services, and data/analytics education. 
\begin{itemize}
\item \emph{Data/analytics design} includes the invention of new methods and ways of designing and producing digital and data products, services, business models, engagement models, communication models, pricing modeling, economic forms, value-added data products/services, decision support systems, automation systems and tools;  
\item \emph{Data/analytics content} includes acquiring, producing, maintaining, publicizing, disseminating, recommending and presenting data-centered content through online, mobile, social media platforms and other channels;  
\item \emph{Data/analytics software} refers to the creation of  software, platforms, architectures, services, tools, systems and applications that acquire, organize, manage, analyze, visualize, use and present data for specific business and scientific purposes, and provide quality assurance to support these  aspects;   
\item \emph{Data/analytics infrastructure} relates to creating infrastructure and devices for data storage, backup, server revenue, data centers, data management and storage, cloud, distributed and parallel computing infrastructure, high performance computing infrastructure, networking, communications, and security;
\item \emph{Data/analytics services} focus on providing strategic and tactical thinking leadership, technical and practical consulting services, problem-oriented solutions and applications, outsourcing, and specific services for data auditing and quality enhancement, data collection, extraction, transformation and loading, recommendation, data center/infrastructure hosting, data analytics and more; 
\item \emph{Data/analytics education} enables the building of corporate competency  and training, as well as offering online/offline/degree-based courses, workshops, materials and services that will allow the gaps in the supply of qualified data professionals to be filled, thus contributing to  building and enhancing the community of this discipline.  
\end{itemize}

The above six core data/analytics areas will see the growth of new data business in terms of the following core aspects and procedures: data storage and management, understanding, processing, optimization, value-added opportunities, transport and communication, servicing and decision support. 

The respective core data/analytics sectors and core procedures can be developed in any data-related sectors, especially data-intensive domains and sectors such as telecommunication, government, finance, banking, capital markets, lifestyle and education. Core business including manufacturing and living business will see increased opportunities for better collection, management and use of data. Analytics will be undertaken to improve productivity, effectiveness, and efficiency, and create new value-added  growth in the economy.  

Interestingly, as we have seen, the data economy occupies the top 10 largest capital entities. The data industry continues to create new business models, products, services, operationalization modes, and workforce models. Data economy will further change the way we live, work, learn and are entertained, as new facilities and environments are created in which data plays a critical role.

\subsection{Data Services}
\label{subsec:service}
Data services form part of the whole landscape of data and analytics which, as noted above, is changing every aspect of the way we live. Data services can be differentiated  from traditional services by the fact that they are not traditional physical material- or energy-oriented services. 
\begin{itemize}
\item Data services act as the core business rather than the  auxiliary business of an economy;
\item Data-driven production and decision-making emerges as the core function  in large organizations for complex decision-making and strategic planning, rather than adjunct facilities;  
\item Data services are online, mobile and socially based, embedded in our activities and agenda;
\item Data business is global and 24/7, offered at any place at any time on demand or in a supply-driven mode;
\item The provision of data services  does not require traditional production elements such as intensive workshops, factories and office facilities;
\item Data-driven services offer real-time public service data management, high performance processing, analytics and decision-making;
\item Data-driven services support full life-cycle analysis, from descriptive, predictive and prescriptive analytics for the prediction, detection,  and prevention of risk, to innovation and optimization; 
\item Data-analytical services are intelligent or can enhance the intelligence of generous data and information services;  
\item Data services enable cross-media, cross-source and cross-organization innovation and practice; and
\item Data services demonstrate significant savings and efficiency improvement through the delivery of actionable knowledge/insights. 
\end{itemize}

Some typical data services delivered through analytics for both core business and new economy are listed below as examples: 
\begin{itemize}
\item \emph{Credit scoring}: to establish the credit worthiness of a customer requesting a loan;
\item \emph{Fraud detection}: to identify fraudulent transactions and suspicious behavior;
\item \emph{Healthcare}: to detect over-service, under-service, fraud, and events like epidemics;
\item \emph{Insurance}: to detect fraudulent claims and assess risk; 
\item \emph{Manufacturing process analysis}: to identify the causes of manufacturing problems and to optimize the processes;
\item \emph{Marketing and sales}: to identify potential customers and establish the effectiveness of campaigns;
\item \emph{Portfolio trading}: to optimize a portfolio of financial instruments by maximizing returns and minimizing risk;
\item \emph{Surveillance}: to detect intrusion, objects, persons and linkages from multi-sensor data and remote sensing;
\item \emph{Understanding customer behaviors}: to model churn, affinities, propensities, and next best actions on intervention behaviors;
\item \emph{Web analytics}: to model user preferences from data to devise and provide personalized and targeted services.
\end{itemize}

A major challenge and increasing need  in the data industry is to provide global or Internet-based data services for a collection of organizations, such as multi-national companies and whole-of-government. Such services need to
\begin{itemize}
\item Define global data analytics objectives and benefits;
\item Support good data governance, security, privacy and accountability to enable smarter data use and sharing;
\item Support data matching and sharing in the context of cross-organizational, cross-platform, cross-format and cross-analytical goals;
\item Prepare global and organization-specific local/departmental data;
\item Foster global and local analytics capabilities, capacity and competency;
\item Enable sharing and collaboration in data and analytics skills, infrastructure, tools, techniques and outcomes;
\item Support crowdsourcing, collaborative and parallel analytic tasks and analytic workflow management;
\item Support analytic capability and package sharing;
\item Support data and data software versioning management and control \cite{Bhardwaj14} at a global and collaborative level;
\item Support the visualization and dissemination of outcomes to targeted audiences and in personalized preferences.
\end{itemize}

Data-driven industry and service are forming new trends in data science for business, for instance:
\begin{itemize}
\item Advanced analytics is no longer just for analysts \cite{Ghodke15CIOL}; dummy analytics is becoming the default setting of management and operational systems;
\item Cloud data management, storage and cloud-based analytics are gaining popularity  \cite{Ghodke15CIOL} and are replacing traditional management information systems, business support systems, and operational support systems; 
\item Data science on scale from multiple sources of data is becoming feasible; Internet-based services are a strongly growing area of the new economy;
\item Analytics as a service is becoming feasible with appropriate social issue management, as  analytics becomes a reality everywhere and is embedded in business, mobile, social and online services;  
\item Visual analytics is becoming a common language;
\item Data services can be mixed with virtual reality and  presented in a way that combines physical and virtual worlds, resources and intelligence;
\item Services on matched and mixed data are streamlined into a  one-world process, with both local and global objectives addressed. 
\end{itemize}

\section{Data Education: Capabilities and Competency}
\label{sec:education}
Data innovation and economy is dependent on the corresponding data and analytics capabilities and competencies and the ability to handle related social issues. These  are weak areas, and there are significant gaps in the current body of knowledge, organizational maturity \cite{Paulk93,Crowston11}, education and training. The requirement is to ``think with data'', ``manage data'', ``compute with data'', ``mine on data'', ``communicate with data'', ``deliver with data'', and ``take action on data'' \cite{Cao16np}. This section discusses these important matters.  

More and more industry and government organizations recognize the value of data for decision-making and have set up general and specific data scientist roles to support data science and engineering, e.g., Chief Data Officer, Chief Analytics Officer, data modelers and data miners, in addition to data engineers and business analysts.   

\subsection{Data Scientists in a Sexy Profession}
\label{subsec:ds-profession}

The role of the data scientist was recognized 10 years ago, and it has become a sexy profession in the job market. The next-generation data scientists will  be mostly welcomed in the increasingly important data economy and data-to-decision.

In 2004, Dr Usama Fayyad was appointed as the Chief Data Officer of Yahoo, which opened the door to a new career possibility: the \emph{data science professional} \cite{Manieri15-1,Harris13} or more specifically \emph{data scientist}, for those people whose role very much centers on data. In 2015, the White House appointed the first U.S. Chief Data Scientist \cite{whcds}. This role ``will shape policies and practices to help the U.S. remain a leader in technology and innovation, foster partnerships to help responsibly maximize the nation's return on its investment in data, and help to recruit and retain the best minds in data science to join us in serving the public.''  \cite{whcds}.

Today, the role of data scientist \cite{Patil11} is regarded as ``the sexiest job of the 21st century'' \cite{Davenport12}. It is reported that data scientists earn much higher salaries than those in  other data-related jobs, with a median salary of US\$120k for data scientists and US\$160k for managers, according to the 2014 Burtchworks survey \cite{Burtch14}. This is attributed to the fact that 88\% of respondents in this survey have at least a Master's degree, while 46\% also hold a Doctorate compared to only 20\% of other Big Data professionals. In the 2015 O'Reilly survey \cite{King15}, 23\% were found to hold a doctorate, while another 44\% had a Master's. The  median annual base salary of this survey sample was US\$91,000 globally, and among US respondents was US\$104,000, compared to US\$150k for ``upper management'' (higher than project and product managers). 

\subsection{What Does a Data Scientist Do}
\label{subsec:ds-responsiblities}

So, what are the roles and responsibilities of data scientists? Here we summarize the findings  from several documents on government initiatives:
\begin{itemize}
\item The US National Science Board defines data scientists as ``the information and computer scientists, database and software engineers and programmers, disciplinary experts, curators and expert annotators, librarians, archivists, and others, who are crucial to the successful management of a digital data collection.'' \cite{NSB05}
\item In a report from the US Committee on Science of the National Science and Technology Council, data scientists are defined as ``Scientists who come from information or computer science backgrounds but learn a subject area and may become scientific data curators in disciplines and advance the art of data science. Focus on all parts of the data life cycle.'' \cite{CSNSTC09}
\item The Joint Information Systems Committee defines data scientists as ``people who work where the research is carried out, or, in the case of data centre personnel, in close collaboration with the creators of the data and may be involved in creative inquiry and analysis, enabling others to work with digital data, and developments in data base technology.'' \cite{jisc08}
\end{itemize}

In business, immense interest has been expressed by multinational vendors, social media and online communities, and information providers, such as IBM \cite{ibmds}, LinkedIn \cite{linkedinds}, KDnuggets \cite{kdnuggets}, Facebook \cite{FBdata} and SIAM \cite{siamjobs} about the roles and responsibilities of data scientists and what makes a good data scientist. For instance, a data scientist metromap was created in Chandrasekaran \cite{Chandrasekaran13}. The metromap covers 10 areas and domains: Fundamentals, Statistics, Programming, Machine Learning, Text Mining/Natural Language Processing, Data Visualization, Big Data, Data Ingestion, Data Munging and Toolbox. Each area/domain is represented as a ``metro line'', with the stations depicting the topics to be learned and understood in a progressive fashion. In addition, INFORMS summarizes the following seven job tasks for data scientists \cite{INFORMS14c}: Business Problem (Question) Framing, Analytics Problem Framing,  Data, Methodology (Approach) Selection, Model Building, Deployment, and Model Lifecycle Management. 

An increasing number of academic and research institutions are working on defining the certification and accreditation of next-generation data scientists. This is reflected in general and domain-specific data science curricula for Masters and PhD qualifications, such as a PhD in Analytics \cite{UTSA} and Master's degree in SCM predictive analytics \cite{INFORMS14c}.

Without the loss of generality, typical domain-free and problem-neutral responsibilities and requirements for jobs announced in social media channels (e.g., Google Groups, Facebook and LinkedIn) and what we have experienced in the past 15 years in large governmental and business organizations can be summarized as follows:
\begin{itemize}
\item Learn the business problem domain, talk to business experts and decision-makers to understand the business objectives, requirements and preferences, issues and constraints facing an organization; understand the organizational maturity; identify, specify and define the problems, boundaries and environment, as well as the challenges;  generate business understanding reports;  
\item Identify and specify social and ethical issues such as privacy and security; develop ethical reasoning plans to address social and ethical issues;
\item Understand data characteristics and complexities; identify the problems and constraints of the data; develop a data understanding report; specify and scope analytical goals and milestones by developing respective project plans to set up an agenda and create governance and management plans; 
\item Set up engineering and analytical processes corresponding to analytical goals for turning business and data into information, turning information into insight, and turning insight into business decision-making actions by developing technical plans for the discovery, upgrade and deployment of relevant data intelligence;
\item Transform business problems into analytical tasks, and conduct advanced analytics by developing corresponding techniques, models, methods, algorithms, tools and systems, experimental design and evaluation of data science, generating better practices experience, performing descriptive, predictive and prescriptive analytics, conducting survey research, and supporting visualization and presentation;
\item Based on the understanding of data characteristics and complexities, extract, analyze, construct, mine and select discriminative features, constantly optimize and innovate new variables for best possible problem representation and modeling; when necessary, conduct data quality enhancement \cite{Hazena14};
\item Combine analytical, statistical, algorithmic, engineering and technical skills to mine relevant data by involving contextual information; invent novel and effective models, and constantly improve modeling techniques to optimize and boost model performance and seek to achieve best practice;
\item Maintain, manage and refine projects and milestones, and their processes, deliverables, evaluation, risk and reporting to build active, lifecycle management; 
\item Develop corresponding services, solutions and products or modules to  feed into a system package on top of user-specified programming languages, frameworks and infrastructure, or open source tools and frameworks;
\item Maintain the privacy, security and veracity of data and deliverables;
\item Engage in frequent client interaction during the whole lifecycle; tell clear and concise stories and draw simple conclusions from complex data or algorithms; provide clients with situational analyses and deep insights into areas requiring  improvement; translate into business improving actions in the  final deployment;
\item  Write coherent reports and make presentations to specialists and non-specialists; present executive summaries with precise and evidence-based recommendations and risk management strategies, especially for decision-makers and business owners.
\end{itemize}

\subsection{What Makes a Good Data Scientist}
\label{subsec:ds-qualifications}

To satisfy the above position requirements, data scientist candidates need to have certain qualifications in addition to the analytic skills that are the foundation of this role. These qualifications and abilities include:
\begin{itemize}
\item Thinking, mindset and ability to think analytically, creatively, critically and inquisitively;
\item Methodologies and knowledge of complex systems and approaches for conducting both top-down and bottom-up problem-solving;
\item Master's or PhD degree in computer science, statistics, mathematics, analytics, data science, informatics, engineering, physics, operations research, pattern recognition, artificial intelligence, visualization, information retrieval or related fields;
\item A deep understanding of common statistics, data mining and machine learning methodologies and models;
\item Ability to implement, maintain, and troubleshoot big data infrastructure, such as cloud computing, high performance computing infrastructure, distributed processing paradigms, stream processing and databases;
\item Knowledge of human-computer interactions, visualization and knowledge representation and management;
\item Background in software engineering (including systems design and analysis), quality assurance;
\item Experience working with large datasets, and mixed data types and sources in a networked and distributed environment;
\item Experience in data extraction and processing, feature understanding and relation analysis;
\item Active interest and knowledge in multi-disciplinary and trans-disciplinary studies and methods in scientific, technical, and social and life sciences;
\item Substantial experience with state-of-the-art analytics-oriented scripting, data structures, programming languages, and development platforms in a Linux, cloud or distributed environment;
\item Theoretical background and domain knowledge for the evaluation of the technical and business merits of analytic findings;
\item Excellent written and verbal communication \cite{Matsudaira15} and organizational skills, ability to write and edit analytical materials and reports for different audiences, and capacity to transform analytical concepts and outcomes into business-friendly interpretations; ability to communicate actionable insights to non-technical audiences, and experience in data-driven decision making.
\end{itemize}

While there is significant role overlap between \textit{data scientists} and \textit{business intelligence (BI) professionals} \cite{SAS13}, different research works show that data science professionals are generally much more data and technology-savvy rather than business-oriented, with most holding a Master's or PhD degree in statistics or computer science. 

An EMC data science community survey \cite{emcsurvey11} shows that (1) data scientists can open up new possibilities; (2) compared to 37\% of BI professionals trained in business, 24\% of  data science professionals are in computer science, 17\% are in engineering and 11\% are in hard science;  (3) compared to BI toolkits, data science toolkits are more technically sophisticated and more diversified; (4) the number of data scientists undertaking data experiments is almost double that of BI professionals; (5) data science professionals more frequently interact with diverse technical and business roles in an organization (such as data scientists, strategic planners, statisticians, marketing staff, sales people, graphic designers, business management and IT administration,  programmers, and HR personnel) than BI professionals; (6) compared to working on normal data, big data manipulators tend to tackle more sophisticated data complexities; and (7) data science professionals spend almost double the time they spend on normal data on big data manipulation (e.g., data parsing, organization, mining, algorithms, visualization, story-telling, dynamics, and decisions).  

As a data-centric expert, a good data scientist is also expected to know the underlying domain well. Without an in-depth understanding of the domain, the actionability of the data deliverables and products by data scientists may be low. However, a data scientist is no substitute for domain experts in complex data science problem solving \cite{Cao16ds}. Similar to any other disciplinary specialists, data scientists work more effectively by collaborating with domain-specific specialists and subject matter experts to achieve broader impact. This is similar to the requirements of domain-driven, actionable knowledge discovery \cite{dddm10}.

\subsection{Tools for Data Scientists}
\label{subsec:ds-tools}

In the above sections, the respective responsibilities and qualifications of data scientists have been discussed. In Section \ref{sec:innovation}, relevant research challenges and issues have been listed. To support the data services listed in Section \ref{subsec:service}, the corresponding roles and qualifications discussed in Sections \ref{subsec:ds-responsiblities} and \ref{subsec:ds-qualifications} are necessary. 

In this section, we discuss tools that may be used by data scientists to address the above aspects. Tools are categorized in terms of cloud infrastructure, data and application integration, data preparation and processing, analytics, visualization, programming, master data management,  high performance processing, business intelligence reporting, and project management. A data scientist may use one or more of these tools on demand for data science problem-solving.
\begin{itemize}
\item Cloud infrastructure: Such as Apache Hadoop, Spark, Cloudera, Amazon Web Services, Unix shell/awk/gawk, 1010data, Hortonworks, Pivotal, and MapR. Most traditional IT vendors have migrated their services and platforms to support cloud. 
\item Data/application integration: Including Ab Initio, Informatica, IBM InfoSphere DataStage, Oracle Data Integrator, SAP Data Integrator, Apatar, CloverETL, Information Builders, Jitterbit, Adeptia Integration Suite, DMExpress Syncsort, Pentaho Data Integration, and Talend \cite{solutionsreview-datainteg}.
\item Master data management: Typical software and platforms include IBM InfoSphere Master Data Management Server, Informatica MDM, Microsoft Master Data Services, Oracle	Master Data Management Suite, SAPNetWeaver Master Data Management tool, Teradata Warehousing, TIBCO MDM, Talend MDM, Black Watch Data. 
\item Data preparation and processing: In Today  \cite{predictiveanalyticstoday-dataprep}, 29 data preparation tools and platforms were listed, such as Platfora, Paxata, Teradata Loom, IBM SPSS, Informatica Rev, Omniscope, Alpine Chorus, Knime, and Wrangler Enterprise and Wrangler.
\item Analytics: In addition to well-recognized commercial tools including 	SAS Enterprise Miner, IBM SPSS Modeler and SPSS Statistics, MatLab and Rapidminer \cite{Rapidminer}, many new tools have been created, such as DataRobot \cite{DataRobot}, BigML \cite{BigML}, MLBase \cite{MLBase}, and APIs including Google Cloud Prediction API \cite{Google-api}.
\item Visualization: Many free and commercial software are listed in KDnuggets \cite{kdnuggets-vis} for visualization, such as Interactive Data Language, IRIS Explorer, Miner3D, NETMAP, Panopticon, ScienceGL, Quadrigram, and VisuMap.    
\item Programming: In addition to the main languages R, SAS, SQL, Python and Java, many others are used for analytics, including Scala, JavaScript, .net, NodeJS, Obj-C, PHP, Ruby, and Go \cite{informationweek-proglang}.
\item High performance processing: In Wikipedia  \cite{wikipedia-hpc}, about 40 computer cluster software are listed and compared in terms of their technical performance, such as Stacki, Kubernetes, Moab Cluster Suite, and Platform Cluster Manager.
\item Business intelligence reporting: There are many reporting tools available \cite{Reporting-tools,Wikipedia-reporting}, typical of which are Excel, IBM Cognos, MicroStrategy, SAS Business Intelligence, and SAP Crystal Reports.
\item Project management: In Capterra \cite{Capterra-pm}, more than 500 software and tools were listed for project management, including Microsoft Project, Atlassian, Podio, Wrike, Basecamp, and Teamwork.
\item Social network analysis: In Desale \cite{30-snavt}, 30 tools were listed for SNA and visualization, such as Centrifuge, Commetrix, Cuttlefish, Cytoscape, EgoNet, InFlow, JUNG, Keynetiq, NetMiner, Network Workbench, NodeXL, and SocNetV (Social Networks Visualizer). 
\item Other tools: Increasing numbers of tools have been developed and are under development for domain-specific and problem-specific data science, such as Alteryx and Tableau for tablets; SuggestGrid and Mortar Recommendation Engine for recommender systems \cite{Github-rs}; OptumHealth, Verisk Analytics, MedeAnalytics, McKesson and Truven Health Analytics \cite{Technavio-healthana} for healthcare analytics; BLAST, EMBOSS, Staden, THREADER, PHD and RasMol for bioinformatics.  
\end{itemize}

\section{The Future of Data Science}
\label{sec:dsfuture}
There is continuing debate about how data science will evolve in the next 50 years and what it will ultimately look like. With the joint efforts to be made by the entire scientific community, data science will build its systematic scientific foundations, disciplinary structure, theoretical systems, technological families, and engineering tool sets as an independent science.  

The last 50 years since the proposal of the concept ``data science'' has contributed to the progressive and now widespread acceptance of the need for a new science and its initial conceptualization through its transition and transformation from statistics to the merger with existing disciplines and fields. The next 50 years of data science will extend beyond statistics to identify, discover, explore, and define  specific foundational scientific problems and grand challenges. It will build a systematic family of scientific methodologies and methods and self-contained disciplinary systems and curricula that are not merely a relabeled 'salad' created by mixing existing disciplinary components.  

Based on the understanding of the intrinsic challenges and nature of data science \cite{Cao16ds,Cao16np}, the development of data science may seek to:
\begin{itemize}
\item Design and develop \textit{data brain} that can autonomously mimic human brain working mechanisms while recognize, understand, analyze and learn data and environment, infer and reason about knowledge and insight, and correspondingly decide actions; 
\item Deepen our \textit{understanding of data invisibility}  (i.e., \textit{invisible data characteristics, complexities, intelligence and value}),
in particular, to understand their X-complexities and X-intelligence (see Cao and Fayyad \cite{Cao16ds}). The exploration of what we do not know about what we do not know will strengthen our understanding of the capabilities, limitations, and future directions of data science.
\item Broaden conceptual, theoretical and technological systems for data science by enabling cross-disciplinary and trans-disciplinary research, innovation and education. This will address existing issues such as the variations in statistics hypotheses and will discover and propose problems that are currently invisible to broad science or specific fields;
\item Invent \textit{new data representation capabilities}, including designs, structures, schemas and algorithms to make invisible data complexities and unknown characteristics in complex data 
more visible and explicit, and more easily understood or explored;
\item Design \textit{new storage, access and management mechanisms}, including memory, disk and cloud-based mechanisms, to enable the acquisition, storage, access, sampling, and management of richer characteristics and properties in the physical world that have been simplified and filtered by existing systems, and to support scalable, transparent, flexible, interpretable and personalized data manipulation and analytics in real time;  
\item Create \textit{new analytical and learning capabilities}, including original mathematical, statistical and analytical theories, algorithms and models, to disclose the unknown knowledge in unknown space;
\item Build \textit{new intelligent systems and services}, including corporate and Internet-based collaborative platforms and services, to support the automated, or human-data-cooperative, collaborative and collective exploration of invisible and unknown challenges in unknown space;
\item Train \textit{the next-generation data scientists and data professionals} who are qualified for data science problem-solving, with data literacy, thinking, competency, consciousness, curiosity, communication and cognitive intelligence, to work on the above data science agenda;
\item Assure cross-domain and trans-disciplinary cooperation, collaborations and alliance in complex data science problem-solving. This requires the education of competent data scientists who are multi-disciplinary experts, as well as collaboration between data scientists and domain-specific experts; and
\item Discover and invent \textit{data power} as yet unknown to current understanding and imagination, such as new data economy, mobile applications, social applications, and data-driven business. 
\end{itemize}

\section{Conclusions}
\label{sec:concl}
Data science, big data and advanced analytics have been increasingly recognized as major driving forces for next-generation innovation, economy, and education. Although they are at an early stage of development, strategic discussions about the big picture, trends, major challenges, future directions, and prospects are critical for the healthy development of the field and the community. The purpose of this article has been to share an overview of the conceptualization, development, observations and thinking about the age of data science initiatives, research, innovation, industrialization, profession, competency and education.

We are witnessing a highly evolving data world that seamlessly connects to our daily life, work, learning, economy and entertainment. 
New efforts are increasingly being made by government, industry, academia and even private institutions on ways to convert data for decision-making, and promote the research and development of data science and analytics. The next generation of data science, encompassing a broad range of disciplines, science and economy, relies heavily on the strategic planning and visionary actions that will be undertaken in prioritized data research areas and start-ups. Without any doubt, today's questions such as ``why do we need data science'' will be replaced by a family of scientific theories and tools to address the visible grand challenges and significant problems facing tomorrow's big data, science, business, society, and the economy. We will be greatly amazed by the surprising developments and potential changes that will take place in the next 50 years.

%

\begin{acks}
This work is partially sponsored by the Australian Research Council Discovery Grant (DP130102691).
\end{acks}

\bibliographystyle{ACM-Reference-Format-Journals}

\bibliography{dsa-cusr-24-final}


\begin{thebibliography}{00}


\ifx \showCODEN    \undefined \def \showCODEN     #1{\unskip}     \fi
\ifx \showDOI      \undefined \def \showDOI       #1{{\tt DOI:}\penalty0{#1}\ }
  \fi
\ifx \showISBNx    \undefined \def \showISBNx     #1{\unskip}     \fi
\ifx \showISBNxiii \undefined \def \showISBNxiii  #1{\unskip}     \fi
\ifx \showISSN     \undefined \def \showISSN      #1{\unskip}     \fi
\ifx \showLCCN     \undefined \def \showLCCN      #1{\unskip}     \fi
\ifx \shownote     \undefined \def \shownote      #1{#1}          \fi
\ifx \showarticletitle \undefined \def \showarticletitle #1{#1}   \fi
\ifx \showURL      \undefined \def \showURL       #1{#1}          \fi

\bibitem[\protect\citeauthoryear{ACEMS}{ACEMS}{2014}]%
        {ACEMS}
{ACEMS}. 2014.
\newblock The Australian Research Council (ARC) Centre of Excellence for
  Mathematical and Statistical Frontiers.
\newblock   (2014).
\newblock
\newblock
\shownote{Available at \url{acems.org.au/}.}


\bibitem[\protect\citeauthoryear{Agarwal and Dhar}{Agarwal and Dhar}{2014}]%
        {Agarwal14}
{Ritu Agarwal} {and} {Vasant Dhar}. 2014.
\newblock \showarticletitle{{Editorial-Big} Data, Data Science, and Analytics:
  The Opportunity and Challenge for {IS} Research}.
\newblock {\em Information Systems Research\/} {25}, 3 (2014), 443--448.
\newblock


\bibitem[\protect\citeauthoryear{Agency}{Agency}{2016}]%
        {13th5YP}
{Xinhua~News Agency}. 2016.
\newblock The 13th Five-Year Plan for the National Economic and Social
  Development of the People's Republic of China.
\newblock   (2016).
\newblock
\newblock
\shownote{Available at
  \url{http://news.xinhuanet.com/politics/2016lh/2016-03/17/c_1118366322.htm}.}


\bibitem[\protect\citeauthoryear{AGIMO}{AGIMO}{2013}]%
        {ABDS13}
{AGIMO}. 2013.
\newblock {AGIMO} Big Data Strategy - Issues Paper.
\newblock   (2013).
\newblock
\newblock
\shownote{Available at
  \url{www.finance.gov.au/files/2013/03/Big-Data-Strategy-Issues-Paper1.pdf}.}


\bibitem[\protect\citeauthoryear{Anderson, Bowring, McCauley, Pothering, and
  Starr}{Anderson et~al\mbox{.}}{2014}]%
        {Anderson14-1}
{Paul~E. Anderson}, {James~F. Bowring}, {Renée McCauley}, {George Pothering},
  {and} {Christopher~W. Starr}. 2014.
\newblock \showarticletitle{An undergraduate degree in data science: Curriculum
  and a decade of implementation experience}. In {\em Computer Science
  Education: Proceedings of the 45th ACM Technical Symposium (SIGCSE'14)}.
  145--150.
\newblock


\bibitem[\protect\citeauthoryear{ASA}{ASA}{2015}]%
        {asanews15}
{ASA}. 2015.
\newblock {ASA} views on data science.
\newblock   (2015).
\newblock
\newblock
\shownote{Available at
  \url{http://magazine.amstat.org/?s=data+science&x=0&y=0}.}


\bibitem[\protect\citeauthoryear{AU}{AU}{1990}]%
        {AUdm90}
{AU}. 1990.
\newblock Data-matching Program.
\newblock   (1990).
\newblock
\newblock
\shownote{Available at \url{http://www.comlaw.gov.au/Series/C2004A04095}.}


\bibitem[\protect\citeauthoryear{AU}{AU}{2010}]%
        {AUopengov10}
{AU}. 2010.
\newblock Declaration of Open Government.
\newblock   (2010).
\newblock
\newblock
\shownote{Available at
  \url{http://agimo.gov.au/2010/07/16/declaration-of-open-government/}.}


\bibitem[\protect\citeauthoryear{AU}{AU}{2013}]%
        {AUopengov13}
{AU}. 2013.
\newblock Attorney-General's Department.
\newblock   (2013).
\newblock
\newblock
\shownote{Available at
  \url{http://www.attorneygeneral.gov.au/Mediareleases/Pages/2013/Second\%20quarter/22May2013-AustraliajoinsOpenGovernmentPartnership.aspx}.}


\bibitem[\protect\citeauthoryear{AU}{AU}{2016}]%
        {AUbd}
{AU}. 2016.
\newblock Australia Big Data.
\newblock   (2016).
\newblock
\newblock
\shownote{Available at \url{http://www.finance.gov.au/big-data/}.}


\bibitem[\protect\citeauthoryear{Ayankoya, Calitz, and Greyling}{Ayankoya
  et~al\mbox{.}}{2014}]%
        {Ayankoya14}
{K. Ayankoya}, {A. Calitz}, {and} {J. Greyling}. 2014.
\newblock \showarticletitle{Intrinsic relations between data science, big data,
  business analytics and datafication}.
\newblock {\em ACM International Conference Proceeding Series\/}  {28} (2014),
  192--198.
\newblock


\bibitem[\protect\citeauthoryear{Bailer, Hoerl, Madigan, Montaquila, and
  Wright}{Bailer et~al\mbox{.}}{2012}]%
        {Bailer12}
{J. Bailer}, {R. Hoerl}, {D. Madigan}, {J. Montaquila}, {and} {T. Wright}.
  2012.
\newblock \showarticletitle{Report of the {ASA} Workgroup on Master's Degrees}.
\newblock  (2012).
\newblock


\bibitem[\protect\citeauthoryear{Baumer}{Baumer}{2015}]%
        {Baumer15}
{Ben Baumer}. 2015.
\newblock \showarticletitle{A data science course for undergraduates: Thinking
  with data}.
\newblock {\em The American Statistician\/} {69}, 4 (2015), 334--342.
\newblock


\bibitem[\protect\citeauthoryear{BDL}{BDL}{2016a}]%
        {BDL}
{BDL}. 2016a.
\newblock Big Data Landscape.
\newblock   (2016).
\newblock
\newblock
\shownote{Available at \url{www.bigdatalandscape.com}.}


\bibitem[\protect\citeauthoryear{BDL}{BDL}{2016b}]%
        {bdl16}
{BDL}. 2016b.
\newblock Big Data Landscape 2016 (Version 3.0).
\newblock   (2016).
\newblock
\newblock
\shownote{Available at
  \url{http://mattturck.com/2016/02/01/big-data-landscape/}.}


\bibitem[\protect\citeauthoryear{Beyer and Laney}{Beyer and Laney}{2012}]%
        {Beyer12}
{Mark~A. Beyer} {and} {Douglas Laney}. 2012.
\newblock The Importance of `Big Data': A Definition.
\newblock   (2012).
\newblock
\newblock
\shownote{Available at \url{https://www.gartner.com/doc/2057415}.}


\bibitem[\protect\citeauthoryear{Bhardwaj, Bhattacherjee, Chavan, Deshp,
  Elmore, Madden, and Parameswaran}{Bhardwaj et~al\mbox{.}}{2015}]%
        {Bhardwaj14}
{Anant Bhardwaj}, {Souvik Bhattacherjee}, {Amit Chavan}, {Amol Deshp},
  {Aaron~J. Elmore}, {Samuel Madden}, {and} {Aditya Parameswaran}. 2015.
\newblock \showarticletitle{Datahub: Collaborative data science \& dataset
  version management at scale}. In {\em In CIDR}.
\newblock


\bibitem[\protect\citeauthoryear{BigML}{BigML}{2016}]%
        {BigML}
{BigML}. 2016.
\newblock BigML.
\newblock   (2016).
\newblock
\newblock
\shownote{Available at \url{https://bigml.com/}.}


\bibitem[\protect\citeauthoryear{Borne, Jacoby, Carney, Connolly, Eastman,
  Raddick, Tyson, and Wallin}{Borne et~al\mbox{.}}{2010}]%
        {Borne10}
{Kirk~D. Borne}, {Suzanne Jacoby}, {Karen Carney}, {Andy Connolly}, {Timothy
  Eastman}, {M.~Jordan Raddick}, {J.~A. Tyson}, {and} {John Wallin}. 2010.
\newblock \showarticletitle{The revolution in astronomy education: Data science
  for the masses}.
\newblock  (2010).
\newblock
\newblock
\shownote{Available at \url{http://arxiv.org/pdf/0909.3895v1.pdf}.}


\bibitem[\protect\citeauthoryear{Boyer, Gelman, Schreck, and
  Veeramachaneni}{Boyer et~al\mbox{.}}{2015}]%
        {Boyer15}
{Sebastien Boyer}, {Ben~U. Gelman}, {Benjamin Schreck}, {and} {Kalyan
  Veeramachaneni}. 2015.
\newblock \showarticletitle{Data science foundry for {MOOCs}}. In {\em IEEE
  International Conference on Data Science and Advanced Analytics (DSAA)}.
  1--10.
\newblock


\bibitem[\protect\citeauthoryear{Breiman}{Breiman}{2001}]%
        {Breiman01}
{Leo Breiman}. 2001.
\newblock \showarticletitle{Statistical modeling: The two cultures}.
\newblock {\em Statist. Sci.\/} {16}, 3 (2001), 199--231.
\newblock


\bibitem[\protect\citeauthoryear{Brown}{Brown}{2009}]%
        {Brown}
{Gavin Brown}. 2009.
\newblock Review of Education in Mathematics, Data Science and Quantitative
  Disciplines: Report to the Group of Eight Universities.
\newblock   (2009).
\newblock
\newblock
\shownote{Available at
  \url{https://go8.edu.au/publication/go8-review-education-mathematics-data-science-and-quantitative-disciplines}.}


\bibitem[\protect\citeauthoryear{Burtch}{Burtch}{2014}]%
        {Burtch14}
{Linda Burtch}. 2014.
\newblock The Burtch Works Study: Salaries of Data Scientists.
\newblock   (April 2014).
\newblock
\newblock
\shownote{Available at
  \url{http://www.burtchworks.com/files/2014/07/Burtch-Works-Study_DS_final.pdf}.}


\bibitem[\protect\citeauthoryear{Bussaban and Waraporn}{Bussaban and
  Waraporn}{2015}]%
        {Bussaban15}
{Kanyarat Bussaban} {and} {Phanu Waraporn}. 2015.
\newblock \showarticletitle{Preparing undergraduate students majoring in
  computer science and mathematics with data science perspectives and awareness
  in the age of big data}. In {\em 7th World Conference on Educational
  Sciences}, Vol. 197. 1443--1446.
\newblock


\bibitem[\protect\citeauthoryear{CA}{CA}{2016}]%
        {CAbd}
{CA}. 2016.
\newblock Canada Capitalizing on Big Data.
\newblock   (2016).
\newblock
\newblock
\shownote{Available at
  \url{http://www.sshrc-crsh.gc.ca/news_room-salle_de_presse/latest_news-nouvelles_recentes/big_data_consultation-donnees_massives_consultation-eng.aspx}.}


\bibitem[\protect\citeauthoryear{Cao}{Cao}{2010a}]%
        {Cao10dddm}
{Longbing Cao}. 2010a.
\newblock \showarticletitle{Domain driven data mining: Challenges and
  prospects}.
\newblock {\em IEEE Trans. on Knowledge and Data Engineering\/} {22}, 6 (2010),
  755--769.
\newblock


\bibitem[\protect\citeauthoryear{Cao}{Cao}{2010b}]%
        {Cao10-1}
{Longbing Cao}. 2010b.
\newblock \showarticletitle{In-depth behavior understanding and use: The
  behavior informatics approach}.
\newblock {\em Information Science\/} {180}, 17 (2010), 3067--3085.
\newblock


\bibitem[\protect\citeauthoryear{Cao}{Cao}{2011}]%
        {caoly11}
{Longbing Cao}. 2011.
\newblock Strategic Recommendations on Advanced Data Industry and Services for
  the Yanhuang Science and Technology Park.
\newblock   (2011).
\newblock


\bibitem[\protect\citeauthoryear{Cao}{Cao}{2014}]%
        {noniid14}
{Longbing Cao}. 2014.
\newblock \showarticletitle{Non-IIDness learning in behavioral and social
  data}.
\newblock {\it Comput. J.} {57}, 9 (2014), 1358--1370.
\newblock


\bibitem[\protect\citeauthoryear{Cao}{Cao}{2015a}]%
        {Cao15cl}
{Longbing Cao}. 2015a.
\newblock \showarticletitle{Coupling learning of complex interactions}.
\newblock {\em J. Information Processing and Management\/} {51}, 2 (2015),
  167--186.
\newblock


\bibitem[\protect\citeauthoryear{Cao}{Cao}{2015b}]%
        {Metasynthetic15}
{Longbing Cao}. 2015b.
\newblock {\em Metasynthetic Computing and Engineering of Complex Systems}.
\newblock Springer.
\newblock


\bibitem[\protect\citeauthoryear{Cao}{Cao}{2016a}]%
        {cao16-1}
{Longbing Cao}. 2016a.
\newblock \showarticletitle{Data Science and Analytics: A New Era}.
\newblock {\em International Journal of Data Science and Analytics\/} {1}, 1
  (2016), 1--2.
\newblock


\bibitem[\protect\citeauthoryear{Cao}{Cao}{2016b}]%
        {Cao16ds}
{Longbing Cao}. 2016b.
\newblock Data science: Challenges and directions.
\newblock   (2016).
\newblock
\newblock
\shownote{Technical Report, UTS Advanced Analytics Institute.}


\bibitem[\protect\citeauthoryear{Cao}{Cao}{2016c}]%
        {Cao16np}
{Longbing Cao}. 2016c.
\newblock Data Science: Nature and Pitfalls.
\newblock   (2016).
\newblock
\newblock
\shownote{Technical Report, UTS Advanced Analytics Institute.}


\bibitem[\protect\citeauthoryear{Cao}{Cao}{2016d}]%
        {Cao16pe}
{Longbing Cao}. 2016d.
\newblock Data Science: Profession and Education.
\newblock   (2016).
\newblock
\newblock
\shownote{Technical Report, UTS Advanced Analytics Institute.}


\bibitem[\protect\citeauthoryear{Cao}{Cao}{2017}]%
        {Cao-uds-2017}
{Longbing Cao}. 2017.
\newblock {\em Understand Data Science (to be published)}.
\newblock Springer.
\newblock


\bibitem[\protect\citeauthoryear{Cao and Dai}{Cao and Dai}{2008}]%
        {ocis08}
{Longbing Cao} {and} {Ruwei Dai}. 2008.
\newblock {\em Open Complex Intelligent Systems}.
\newblock Post \& Telecom Press.
\newblock


\bibitem[\protect\citeauthoryear{Cao, Dai, and Zhou}{Cao et~al\mbox{.}}{2009}]%
        {Metasynthesis09}
{Longbing Cao}, {Ruwei Dai}, {and} {Mengchu Zhou}. 2009.
\newblock \showarticletitle{Metasynthesis: {M-Space}, {M-Interaction} and
  {M-Computing} for Open Complex Giant Systems}.
\newblock {\em IEEE Trans. On Systems, Man, and Cybernetics--Part A\/} {39}, 5
  (2009), 1007--1021.
\newblock


\bibitem[\protect\citeauthoryear{Cao and (Eds)}{Cao and (Eds)}{2012}]%
        {Cao12-1}
{Longbing Cao} {and} {Philip S~Yu (Eds)}. 2012.
\newblock {\em Behavior Computing: Modeling, Analysis, Mining and Decision}.
\newblock Springer.
\newblock


\bibitem[\protect\citeauthoryear{Cao, Ou, and Yu}{Cao et~al\mbox{.}}{2012}]%
        {Cao2012cba}
{Longbing Cao}, {Yuming Ou}, {and} {Philip~S Yu}. 2012.
\newblock \showarticletitle{Coupled behavior analysis with applications}.
\newblock {\em IEEE Trans. on Knowledge and Data Engineering\/} {24}, 8 (2012),
  1378--1392.
\newblock


\bibitem[\protect\citeauthoryear{Cao, Yu, Zhang, and Zhao}{Cao
  et~al\mbox{.}}{2010}]%
        {dddm10}
{Longbing Cao}, {Philip~S Yu}, {Chengqi Zhang}, {and} {Yanchang Zhao}. 2010.
\newblock {\em Domain Driven Data Mining}.
\newblock Springer.
\newblock


\bibitem[\protect\citeauthoryear{Capterra}{Capterra}{2016a}]%
        {Capterra-pm}
{Capterra}. 2016a.
\newblock Top Project Management Tools.
\newblock   (2016).
\newblock
\newblock
\shownote{Available at
  \url{http://www.capterra.com/project-management-software/}.}


\bibitem[\protect\citeauthoryear{Capterra}{Capterra}{2016b}]%
        {Reporting-tools}
{Capterra}. 2016b.
\newblock Top Reporting Software Products.
\newblock   (2016).
\newblock
\newblock
\shownote{Available at \url{http://www.capterra.com/reporting-software/}.}


\bibitem[\protect\citeauthoryear{CBDIO}{CBDIO}{2016}]%
        {CBDIO}
{CBDIO}. 2016.
\newblock China Big Data Industrial Observation.
\newblock   (2016).
\newblock
\newblock
\shownote{Available at \url{www.cbdio.com}.}


\bibitem[\protect\citeauthoryear{CCF-BDTF}{CCF-BDTF}{2013}]%
        {CCFBDTF}
{CCF-BDTF}. 2013.
\newblock China Computer Federation Task Force on Big Data.
\newblock   (2013).
\newblock
\newblock
\shownote{Available at \url{http://www.bigdataforum.org.cn/}.}


\bibitem[\protect\citeauthoryear{Chambers}{Chambers}{1993}]%
        {Chambers93}
{John~M Chambers}. 1993.
\newblock \showarticletitle{Greater or lesser statistics: A choice for future
  research}.
\newblock {\em Statistics and Computing\/} {3}, 4 (1993), 182--184.
\newblock


\bibitem[\protect\citeauthoryear{Chandrasekaran}{Chandrasekaran}{2013}]%
        {Chandrasekaran13}
{Swami Chandrasekaran}. 2013.
\newblock Becoming a Data Scientist.
\newblock   (2013).
\newblock
\newblock
\shownote{Available at
  \url{http://nirvacana.com/thoughts/becoming-a-data-scientist/}.}


\bibitem[\protect\citeauthoryear{Chen, Chiang, and Storey}{Chen
  et~al\mbox{.}}{2012}]%
        {Chen12}
{H. Chen}, {R.~H.~L. Chiang}, {and} {V.~C. Storey}. 2012.
\newblock \showarticletitle{Business intelligence and analytics: From Big Data
  to big impact}.
\newblock {\em MIS Quarterly\/} {36}, 4 (2012), 1165--1188.
\newblock


\bibitem[\protect\citeauthoryear{Clancy, Bowles, Gelinas, Androwich, Delaney,
  Matney, Sensmeier, Warren, Welton, and Westra}{Clancy et~al\mbox{.}}{2014}]%
        {Clancy14}
{Thomas~R. Clancy}, {Kathryn~H. Bowles}, {Lillee Gelinas}, {Ida Androwich},
  {Connie Delaney}, {Susan Matney}, {Joyce Sensmeier}, {Judith Warren}, {John
  Welton}, {and} {Bonnie Westra}. 2014.
\newblock \showarticletitle{A call to action: Engage in big data science}.
\newblock {\em Nursing Outlook\/} {62}, 1 (2014), 64--65.
\newblock


\bibitem[\protect\citeauthoryear{Classcentral}{Classcentral}{2016}]%
        {Classcentral}
{Classcentral}. 2016.
\newblock Data Science and Big Data | Free Online Courses.
\newblock   (2016).
\newblock
\newblock
\shownote{Available at
  \url{https://www.class-central.com/subject/data-science}.}


\bibitem[\protect\citeauthoryear{Clay}{Clay}{2013}]%
        {Clay13}
{Kelly Clay}. 2013.
\newblock CES 2013: The Year of The Quantified Self?
\newblock   (2013).
\newblock
\newblock
\shownote{Available at
  \url{http://www.forbes.com/sites/kellyclay/2013/01/06/ces-2013-the-year-of-the-quantified-self/#4cf4d2b55e74}.}


\bibitem[\protect\citeauthoryear{Cleveland}{Cleveland}{2001}]%
        {Cleveland01}
{William~S. Cleveland}. 2001.
\newblock \showarticletitle{Data science: An action plan for expanding the
  technical areas of the field of statistics}.
\newblock {\em International Statistical Review\/} {69}, 1 (2001), 21--26.
\newblock
\showDOI{%
\url{http://dx.doi.org/10.1111/j.1751-5823.2001.tb00477.x}}


\bibitem[\protect\citeauthoryear{CMIST}{CMIST}{2016}]%
        {MIST16}
{CMIST}. 2016.
\newblock China Will Establish A Series of National Labs.
\newblock   (2016).
\newblock
\newblock
\shownote{Available at
  \url{http://news.sciencenet.cn/htmlnews/2016/4/344404.shtm}.}


\bibitem[\protect\citeauthoryear{CNSF}{CNSF}{2015}]%
        {CNSF}
{CNSF}. 2015.
\newblock National Science Foundation China.
\newblock   (2015).
\newblock
\newblock
\shownote{Available at \url{http://www.nsfc.gov.cn/}.}


\bibitem[\protect\citeauthoryear{Commission}{Commission}{2014}]%
        {ECbd14}
{European Commission}. 2014.
\newblock \showarticletitle{Commission urges governments to embrace potential
  of big data}.
\newblock  (2014).
\newblock
\newblock
\shownote{Available at \url{europa.eu/rapid/press-release_IP-14-769_en.htm}.}


\bibitem[\protect\citeauthoryear{Coursera}{Coursera}{2016}]%
        {Coursera}
{Coursera}. 2016.
\newblock Coursera.
\newblock   (2016).
\newblock
\newblock
\shownote{Available at \url{www.coursera.org/data-science}.}


\bibitem[\protect\citeauthoryear{Crowston and Qin}{Crowston and Qin}{2011}]%
        {Crowston11}
{Kevin Crowston} {and} {Jian Qin}. 2011.
\newblock \showarticletitle{A capability maturity model for scientific data
  management: Evidence from the literatute}.
\newblock  {48}, 10 (2011), 1--9.
\newblock


\bibitem[\protect\citeauthoryear{CSC}{CSC}{2012}]%
        {CSCbd}
{CSC}. 2012.
\newblock \showarticletitle{Big Data Universe Beginning to Explode}.
\newblock  (2012).
\newblock
\newblock
\shownote{Available at
  \url{http://www.csc.com/insights/flxwd/78931-big_data_growth_just_beginning_to_explode}.}


\bibitem[\protect\citeauthoryear{CSNSTC}{CSNSTC}{2009}]%
        {CSNSTC09}
{CSNSTC}. 2009.
\newblock \showarticletitle{Harnessing the Power of Digital Data for Science
  and Society}.
\newblock  (2009).
\newblock
\newblock
\shownote{Report of the Interagency Working Group on Digital Data to the
  Committee on Science of the National Science and Technology Council.}


\bibitem[\protect\citeauthoryear{DABS}{DABS}{2016}]%
        {DABS}
{DABS}. 2016.
\newblock Data Analytics Book Series.
\newblock   (2016).
\newblock
\newblock
\shownote{Available at \url{http://www.springer.com/series/15063}.}


\bibitem[\protect\citeauthoryear{DARPA}{DARPA}{2016}]%
        {Xdata}
{DARPA}. 2016.
\newblock {DARPA} Xdata program.
\newblock   (2016).
\newblock
\newblock
\shownote{Available at \url{www.darpa.mil/program/xdata}.}


\bibitem[\protect\citeauthoryear{Data61}{Data61}{2016}]%
        {data61}
{Data61}. 2016.
\newblock Data61.
\newblock   (2016).
\newblock
\newblock
\shownote{Available at \url{https://www.data61.csiro.au/}.}


\bibitem[\protect\citeauthoryear{DataRobot}{DataRobot}{2016}]%
        {DataRobot}
{DataRobot}. 2016.
\newblock DataRobot.
\newblock   (2016).
\newblock
\newblock
\shownote{Available at \url{https://www.datarobot.com/}.}


\bibitem[\protect\citeauthoryear{Datasciences.org}{Datasciences.org}{2005}]%
        {datasciences.org}
{Datasciences.org}. 2005.
\newblock Datasciences.org.
\newblock   (2005).
\newblock
\newblock
\shownote{Available at \url{www.datasciences.org}.}


\bibitem[\protect\citeauthoryear{Davenport and Patil}{Davenport and
  Patil}{2012}]%
        {Davenport12}
{Thomas~H. Davenport} {and} {D.J. Patil}. 2012.
\newblock \showarticletitle{Data scientist: The sexiest job of the 21st
  century}.
\newblock {\em Harvard Business Review\/} (2012), 70--76.
\newblock


\bibitem[\protect\citeauthoryear{Davis}{Davis}{2016}]%
        {informationweek-proglang}
{Jessica Davis}. 2016.
\newblock 10 Programming Languages And Tools Data Scientists Used.
\newblock   (2016).
\newblock
\newblock
\shownote{Available at
  \url{http://www.informationweek.com/devops/programming-languages/10-programming-languages-and-tools-data-scientists-use-now/d/d-id/1326034}.}


\bibitem[\protect\citeauthoryear{Desale}{Desale}{2015}]%
        {30-snavt}
{Devendra Desale}. 2015.
\newblock Top 30 Social Network Analysis and Visualization Tools.
\newblock   (2015).
\newblock
\newblock
\shownote{Available at
  \url{http://www.kdnuggets.com/2015/06/top-30-social-network-analysis-visualization-tools.html}.}


\bibitem[\protect\citeauthoryear{Dhar}{Dhar}{2013}]%
        {Dhar13}
{Vasant Dhar}. 2013.
\newblock \showarticletitle{Data science and prediction}.
\newblock {\it Commun. ACM} {56}, 12 (2013), 64--73.
\newblock


\bibitem[\protect\citeauthoryear{Dierick and Gabbiani}{Dierick and
  Gabbiani}{2015}]%
        {Dierick15}
{Herman~A. Dierick} {and} {Fabrizio Gabbiani}. 2015.
\newblock \showarticletitle{Drosophila neurobiology: No escape from `Big Data'
  science}.
\newblock {\em Current Biology\/} {25}, 14 (2015), 606--608.
\newblock


\bibitem[\protect\citeauthoryear{Diggle}{Diggle}{2015}]%
        {Diggle15}
{Peter~J. Diggle}. 2015.
\newblock \showarticletitle{Statistics: A data science for the 21st century}.
\newblock {\em Journal of the Royal Statistical Society: Series A (Statistics
  in Society)\/} {178}, 4 (2015), 793--813.
\newblock


\bibitem[\protect\citeauthoryear{Donoho}{Donoho}{2015}]%
        {Donoho15}
{David Donoho}. 2015.
\newblock 50 years of Data Science.
\newblock   (2015).
\newblock
\newblock
\shownote{Available at
  \url{http://courses.csail.mit.edu/18.337/2015/docs/50YearsDataScience.pdf}.}


\bibitem[\protect\citeauthoryear{Dorr, Greenberg, Fontana, Przybocki, Bras,
  Ploehn, Aulov, Michel, Golden, and Chang}{Dorr et~al\mbox{.}}{2015}]%
        {Dorr15-2}
{Bonnie~J. Dorr}, {Craig~S. Greenberg}, {Peter Fontana}, {Mark~A. Przybocki},
  {Marion~Le Bras}, {Cathryn~A. Ploehn}, {Oleg Aulov}, {Martial Michel},
  {E.~Jim Golden}, {and} {Wo Chang}. 2015.
\newblock \showarticletitle{The {{NIST}} data science initiative}. In {\em 2015
  IEEE International Conference on Data Science and Advanced Analytics (DSAA)}.
  1--10.
\newblock


\bibitem[\protect\citeauthoryear{DSA}{DSA}{2016}]%
        {Datascienceassn}
{DSA}. 2016.
\newblock Data Science Association.
\newblock   (2016).
\newblock
\newblock
\shownote{Available at \url{http://www.datascienceassn.org/}.}


\bibitem[\protect\citeauthoryear{DSAA}{DSAA}{2014}]%
        {DSAA}
{DSAA}. 2014.
\newblock {IEEE/ACM/ASA} International Conference on Data Science and Advanced
  Analytics.
\newblock   (2014).
\newblock
\newblock
\shownote{Available at \url{www.dsaa.co}.}


\bibitem[\protect\citeauthoryear{DSC}{DSC}{2016a}]%
        {DSC}
{DSC}. 2016a.
\newblock College \& University Data Science Degrees.
\newblock   (2016).
\newblock
\newblock
\shownote{Available at \url{http://datascience.community/colleges (accessed on
  16 April 2016.)}.}


\bibitem[\protect\citeauthoryear{DSC}{DSC}{2016b}]%
        {DSCEU}
{DSC}. 2016b.
\newblock The Data Science Community.
\newblock   (2016).
\newblock
\newblock
\shownote{Available at \url{http://datasciencebe.com/}.}


\bibitem[\protect\citeauthoryear{DSCentral}{DSCentral}{2016}]%
        {Dscentral}
{DSCentral}. 2016.
\newblock Data Science Central.
\newblock   (2016).
\newblock
\newblock
\shownote{Available at \url{http://www.datasciencecentral.com/}.}


\bibitem[\protect\citeauthoryear{DSE}{DSE}{2015}]%
        {DSE}
{DSE}. 2015.
\newblock Data Science and Engineering.
\newblock   (2015).
\newblock
\newblock
\shownote{Available at \url{http://link.springer.com/journal/41019}.}


\bibitem[\protect\citeauthoryear{DSJ}{DSJ}{2014}]%
        {DSJ}
{DSJ}. 2014.
\newblock Data Science Journal.
\newblock   (2014).
\newblock
\newblock
\shownote{Available at \url{datascience.codata.org}.}


\bibitem[\protect\citeauthoryear{DSKD}{DSKD}{2007}]%
        {DSKD07}
{DSKD}. 2007.
\newblock Data Science and Knowledge Discovery Lab, {UTS}.
\newblock   (2007).
\newblock
\newblock
\shownote{Available at
  \url{http://www.uts.edu.au/research-and-teaching/our-research/quantum-computation-and-intelligent-systems/data-sciences-and}.}


\bibitem[\protect\citeauthoryear{Duncan}{Duncan}{2009}]%
        {Duncan09}
{David~Ewing Duncan}. 2009.
\newblock {\em Experimental Man: What One Man's Body Reveals about His Future,
  Your Health, and Our Toxic World}.
\newblock New York: Wiley \& Sons.
\newblock


\bibitem[\protect\citeauthoryear{Edx}{Edx}{2016}]%
        {Edx}
{Edx}. 2016.
\newblock {EDX} Courses.
\newblock   (2016).
\newblock
\newblock
\shownote{Available at
  \url{https://www.edx.org/course?search_query=data+science}.}


\bibitem[\protect\citeauthoryear{EMC}{EMC}{2011}]%
        {emcsurvey11}
{EMC}. 2011.
\newblock \showarticletitle{Data Science Revealed: A {Data-Driven} Glimpse into
  the Burgeoning New Field}.
\newblock  (2011).
\newblock
\newblock
\shownote{Available at
  \url{www.emc.com/collateral/about/news/emc-data-science-study-wp.pdf}.}


\bibitem[\protect\citeauthoryear{EPJDS}{EPJDS}{2012}]%
        {EPJDS}
{EPJDS}. 2012.
\newblock {EPJ} Data Science.
\newblock   (2012).
\newblock
\newblock
\shownote{Available at \url{http://epjdatascience.springeropen.com/}.}


\bibitem[\protect\citeauthoryear{EU}{EU}{2014}]%
        {EUbd}
{EU}. 2014.
\newblock {EU} Towards a Thriving Data-driven Economy.
\newblock   (2014).
\newblock
\newblock
\shownote{Available at
  \url{https://ec.europa.eu/digital-single-market/en/towards-thriving-data-driven-economy}.}


\bibitem[\protect\citeauthoryear{EU-DSA}{EU-DSA}{2016}]%
        {EUdsa}
{EU-DSA}. 2016.
\newblock The European Data Science Academy.
\newblock   (2016).
\newblock
\newblock
\shownote{Available at \url{edsa-project.eu}.}


\bibitem[\protect\citeauthoryear{EU-OD}{EU-OD}{2016}]%
        {EUod}
{EU-OD}. 2016.
\newblock The European Union Open Data Portal.
\newblock   (2016).
\newblock
\newblock
\shownote{Available at \url{https://open-data.europa.eu/}.}


\bibitem[\protect\citeauthoryear{Facebook}{Facebook}{2016}]%
        {FBdata}
{Facebook}. 2016.
\newblock Facebook Data.
\newblock   (2016).
\newblock
\newblock
\shownote{Available at \url{https://www.facebook.com/careers/teams/data/}.}


\bibitem[\protect\citeauthoryear{Faghmous and Kumar}{Faghmous and
  Kumar}{2014}]%
        {Faghmous14}
{James~H. Faghmous} {and} {Vipin Kumar}. 2014.
\newblock \showarticletitle{A big data guide to understanding climate change:
  The case for {theory-guided} data science}.
\newblock {\em Big Data\/} {2}, 3 (2014), 155--163.
\newblock


\bibitem[\protect\citeauthoryear{Fairfielda and Shteina}{Fairfielda and
  Shteina}{2014}]%
        {Fairfielda14}
{Joshua Fairfielda} {and} {Hannah Shteina}. 2014.
\newblock \showarticletitle{Big data, big problems: Emerging issues in the
  ethics of data science and journalism}.
\newblock {\em Journal of Mass Media Ethics\/} {29}, 1 (2014), 38--51.
\newblock


\bibitem[\protect\citeauthoryear{Faris, Kolker, Szalay, Bradlow, Deelman, Feng,
  Qiu, Russell, Stewart, and Kolker}{Faris et~al\mbox{.}}{2011}]%
        {Faris11}
{Jack Faris}, {Evelyne Kolker}, {Alex Szalay}, {Leon Bradlow}, {Ewa Deelman},
  {Wu Feng}, {Judy Qiu}, {Donna Russell}, {Elizabeth Stewart}, {and} {Eugene
  Kolker}. 2011.
\newblock \showarticletitle{Communication and data-intensive science in the
  beginning of the 21st century}.
\newblock {\em A Journal of Integrative Biology\/} {15}, 4 (2011), 213--215.
\newblock


\bibitem[\protect\citeauthoryear{Fawcett}{Fawcett}{2016}]%
        {Fawcett16}
{Tom Fawcett}. 2016.
\newblock \showarticletitle{Mining the quantified self: Personal knowledge
  discovery as a challenge for data science}.
\newblock {\em Big Data\/} (2016), 249--266.
\newblock


\bibitem[\protect\citeauthoryear{Fayyad, Piatetsky-Shapiro, and Smyth}{Fayyad
  et~al\mbox{.}}{1996}]%
        {Fayyad96}
{Usama Fayyad}, {Gregory Piatetsky-Shapiro}, {and} {Padhraic Smyth}. 1996.
\newblock \showarticletitle{From data mining to knowledge discovery in
  databases}.
\newblock {\em AI Magazine\/} {17}, 3 (1996), 37--54.
\newblock


\bibitem[\protect\citeauthoryear{Finzer}{Finzer}{2013}]%
        {Finzer13}
{William Finzer}. 2013.
\newblock \showarticletitle{The data science education dilemma}.
\newblock {\em Technology Innovations in Statistics Education\/} {7}, 2 (2013).
\newblock


\bibitem[\protect\citeauthoryear{Fox, Maini, Rosenbaum, and Wild}{Fox
  et~al\mbox{.}}{2015}]%
        {Fox15}
{Geoffrey Fox}, {Siddharth Maini}, {Howard Rosenbaum}, {and} {David~J. Wild}.
  2015.
\newblock \showarticletitle{Data science and online education}. In {\em 2015
  IEEE 7th International Conference on Cloud Computing Technology and Science
  (CloudCom)}. 582--587.
\newblock


\bibitem[\protect\citeauthoryear{Fox and Hendler}{Fox and Hendler}{2014}]%
        {Fox11}
{P. Fox} {and} {J. Hendler}. 2014.
\newblock \showarticletitle{The science of data science}.
\newblock {\em Big Data\/} {2}, 2 (2014), 68--70.
\newblock


\bibitem[\protect\citeauthoryear{Galetto}{Galetto}{2016}]%
        {Galetto16}
{Molly Galetto}. 2016.
\newblock Top 50 Data Science Resources.
\newblock   (2016).
\newblock
\newblock
\shownote{Available at
  \url{http://www.ngdata.com/top-data-science-resources/?}}


\bibitem[\protect\citeauthoryear{GEO}{GEO}{2016}]%
        {GEOdata}
{GEO}. 2016.
\newblock Gene Expression Omnibus.
\newblock   (2016).
\newblock
\newblock
\shownote{Available at \url{http://www.ncbi.nlm.nih.gov/geo/}.}


\bibitem[\protect\citeauthoryear{Ghodke}{Ghodke}{2015}]%
        {Ghodke15CIOL}
{Deepak Ghodke}. 2015.
\newblock Bye Bye 2015: What lies ahead for BI.
\newblock   (2015).
\newblock
\newblock
\shownote{Available at
  \url{http://www.ciol.com/bye-bye-2015-what-lies-ahead-for-bi/}.}


\bibitem[\protect\citeauthoryear{Github}{Github}{2016a}]%
        {Github}
{Github}. 2016a.
\newblock Data science colleges.
\newblock   (2016).
\newblock
\newblock
\shownote{Available at
  \url{https://github.com/ryanswanstrom/awesome-datascience-colleges}.}


\bibitem[\protect\citeauthoryear{Github}{Github}{2016b}]%
        {Github-rs}
{Github}. 2016b.
\newblock List of Recommender Systems.
\newblock   (2016).
\newblock
\newblock
\shownote{Available at
  \url{https://github.com/grahamjenson/list_of_recommender_systems}.}


\bibitem[\protect\citeauthoryear{Gold, McClarren, and Gaughan}{Gold
  et~al\mbox{.}}{2013}]%
        {Gold13}
{Michael Gold}, {Ryan McClarren}, {and} {Conor Gaughan}. 2013.
\newblock \showarticletitle{The lessons Oscar taught us: Data science and media
  \& entertainment}.
\newblock {\em Big Data\/} {1}, 2 (2013), 105--109.
\newblock


\bibitem[\protect\citeauthoryear{Google}{Google}{2016a}]%
        {Google}
{Google}. 2016a.
\newblock Google Bigquery and Cloud Platform.
\newblock   (2016).
\newblock
\newblock
\shownote{Available at \url{https://cloud.google.com/bigquery/}.}


\bibitem[\protect\citeauthoryear{Google}{Google}{2016b}]%
        {Google-api}
{Google}. 2016b.
\newblock Google Cloud Prediction API.
\newblock   (2016).
\newblock
\newblock
\shownote{Available at \url{https://cloud.google.com/prediction/docs/}.}


\bibitem[\protect\citeauthoryear{Google}{Google}{2016c}]%
        {Googlecb}
{Google}. 2016c.
\newblock Google Online Open Education.
\newblock   (2016).
\newblock
\newblock
\shownote{Available at \url{https://www.google.com/edu/openonline/}.}


\bibitem[\protect\citeauthoryear{Google}{Google}{2016d}]%
        {Googletrends}
{Google}. 2016d.
\newblock Google Trends.
\newblock   (2016).
\newblock
\newblock
\shownote{Available at
  \url{https://www.google.com.au/trends/explore#q=data\%20science\%2C\%20data\%20analytics\%2C\%20big\%20data\%2C\%20data\%20analysis\%2C\%20advanced\%20analytics&cmpt=q&tz=Etc\%2FGMT-11}.}


\bibitem[\protect\citeauthoryear{Google}{Google}{2016e}]%
        {Mobiledata}
{Google}. 2016e.
\newblock Open Mobile Data.
\newblock   (2016).
\newblock
\newblock
\shownote{Available at
  \url{https://console.developers.google.com/storage/browser/openmobiledata_public/}.}


\bibitem[\protect\citeauthoryear{Government}{Government}{2016}]%
        {Beijing2016-20}
{Beijing~Municipal Government}. 2016.
\newblock Beijing Big Data and Cloud Computing Development Action Plan.
\newblock   (2016).
\newblock
\newblock
\shownote{Available at \url{http://zhengwu.beijing.gov.cn/gh/dt/t1445533.htm}.}


\bibitem[\protect\citeauthoryear{Government}{Government}{2015}]%
        {CNbd}
{China Government}. 2015.
\newblock China Big Data.
\newblock   (2015).
\newblock
\newblock
\shownote{Available at
  \url{http://www.gov.cn/zhengce/content/2015-09/05/content_10137.htm}.}


\bibitem[\protect\citeauthoryear{Graham}{Graham}{2012}]%
        {Graham12}
{Matthew~J. Graham}. 2012.
\newblock \showarticletitle{The art of data science}. In {\em Astrostatistics
  and Data Mining, Volume 2 of the series Springer Series in Astrostatistics}.
  47--59.
\newblock


\bibitem[\protect\citeauthoryear{Gray}{Gray}{2007}]%
        {Gray07}
{Jim Gray}. 2007.
\newblock eScience -- A Transformed Scientific Method.
\newblock   (2007).
\newblock
\newblock
\shownote{Available at
  \url{http://research.microsoft.com/en-us/um/people/gray/talks/NRC-CSTB_eScience.ppt}.}


\bibitem[\protect\citeauthoryear{GTD}{GTD}{2016}]%
        {Terrorismdata}
{GTD}. 2016.
\newblock Global Terrorism Database.
\newblock   (2016).
\newblock
\newblock
\shownote{Available at \url{https://www.start.umd.edu/gtd/}.}


\bibitem[\protect\citeauthoryear{Gupta, Cecen, Goyal, Singh, and
  Kalidindi}{Gupta et~al\mbox{.}}{2015}]%
        {Gupta15}
{Akash Gupta}, {Ahmet Cecen}, {Sharad Goyal}, {Amarendra~K. Singh}, {and}
  {Surya~R. Kalidindi}. 2015.
\newblock \showarticletitle{Structure-property linkages using a data science
  approach: Application to a non-metallic inclusion/steel composite system}.
\newblock {\em Acta Mater\/}  {91} (2015), 239--254.
\newblock


\bibitem[\protect\citeauthoryear{Hand}{Hand}{2015}]%
        {Hand15}
{David~J. Hand}. 2015.
\newblock \showarticletitle{Statistics and computing: The genesis of data
  science}.
\newblock {\em Statistics and Computing\/} {25}, 4 (2015), 705--711.
\newblock


\bibitem[\protect\citeauthoryear{Hardin}{Hardin}{2016}]%
        {Hardin47}
{Hardin}. 2016.
\newblock Github.
\newblock   (2016).
\newblock
\newblock
\shownote{Available at \url{hardin47.github.io/DataSciStatsMaterials/}.}


\bibitem[\protect\citeauthoryear{Hardin, Hoerl, Horton, Nolan, Baumer,
  Hall-Holt, Murrell, Peng, Roback, Lang, and Ward}{Hardin
  et~al\mbox{.}}{2015}]%
        {Hardin15}
{J. Hardin}, {R. Hoerl}, {Nicholas~J. Horton}, {D. Nolan}, {B. Baumer}, {O.
  Hall-Holt}, {P. Murrell}, {R. Peng}, {P. Roback}, {D.~Temple Lang}, {and}
  {M.~D. Ward}. 2015.
\newblock \showarticletitle{Data science in statistics curricula: Preparing
  students to ``Think with Data''}.
\newblock {\em The American Statistician\/} {69}, 4 (2015), 343--353.
\newblock


\bibitem[\protect\citeauthoryear{Harris, Murphy, and Vaisman}{Harris
  et~al\mbox{.}}{2013}]%
        {Harris13}
{Harlan Harris}, {Sean Murphy}, {and} {Marck Vaisman}. 2013.
\newblock {\em Analyzing the Analyzers: An Introspective Survey of Data
  Scientists and Their Work}.
\newblock
\newblock
\shownote{O'Reilly Media.}


\bibitem[\protect\citeauthoryear{Hazena, Booneb, Ezellc, and
  Jones-Farmer}{Hazena et~al\mbox{.}}{2014}]%
        {Hazena14}
{Benjamin~T. Hazena}, {Christopher~A. Booneb}, {Jeremy~D. Ezellc}, {and}
  {L.~Allison Jones-Farmer}. 2014.
\newblock \showarticletitle{Data quality for data science, predictive
  analytics, and big data in supply chain management: An introduction to the
  problem and suggestions for research and applications}.
\newblock {\em International Journal of Production Economics\/}  {154} (2014),
  72--80.
\newblock


\bibitem[\protect\citeauthoryear{Hey, Tansley, and (Eds.)}{Hey
  et~al\mbox{.}}{2009}]%
        {4thparadigm}
{Tony Hey}, {Stewart Tansley}, {and} {Kristin~Tolle (Eds.)}. 2009.
\newblock {\em The Fourth Paradigm: {Data-Intensive} Scientific Discovery}.
\newblock
\newblock
\shownote{Available at
  \url{http://research.microsoft.com/en-us/collaboration/fourthparadigm/}.}


\bibitem[\protect\citeauthoryear{Hey and Trefethen}{Hey and Trefethen}{2003}]%
        {Hey03}
{Tony Hey} {and} {Anne Trefethen}. 2003.
\newblock {\em The Data Deluge: An e-Science Perspective}.
\newblock John Wiley \& Sons, Ltd, 809--824.
\newblock
\showISBNx{9780470867167}


\bibitem[\protect\citeauthoryear{HLSG}{HLSG}{2010}]%
        {HLSGwave10}
{HLSG}. 2010.
\newblock \showarticletitle{Final Report of the High Level Expert Group on
  Scientific Data}.
\newblock  (2010).
\newblock
\newblock
\shownote{Available at
  \url{http://ec.europa.eu/information_society/newsroom/cf/document.cfm?action=display&doc_id=707}.}


\bibitem[\protect\citeauthoryear{HLSG}{HLSG}{2014}]%
        {HLSGharest14}
{HLSG}. 2014.
\newblock \showarticletitle{An RDA Europe Report}.
\newblock  (2014).
\newblock
\newblock
\shownote{Available at
  \url{http://www.e-nformation.ro/wp-content/uploads/2014/12/TheDataHarvestReport_-Final.pdf}.}


\bibitem[\protect\citeauthoryear{Horizon}{Horizon}{2014}]%
        {Horizonbd}
{Horizon}. 2014.
\newblock European Commission Horizon 2020 Big Data Private Public Partnership.
\newblock   (2014).
\newblock
\newblock
\shownote{Available at
  \url{http://ec.europa.eu/programmes/horizon2020/en/h2020-section/information-and-communication-technologies}.}


\bibitem[\protect\citeauthoryear{Huber}{Huber}{2011}]%
        {Huber11}
{Peter~J. Huber}. 2011.
\newblock {\em Data Analysis: What Can Be Learned From the Past 50 Years}.
\newblock John Wiley \& Sons.
\newblock
\showISBNx{978-1-118-01064-8}


\bibitem[\protect\citeauthoryear{IASC}{IASC}{1977}]%
        {IASC77}
{IASC}. 1977.
\newblock International Association for Statistical Computing.
\newblock   (1977).
\newblock
\newblock
\shownote{Available at \url{http://www.iasc-isi.org/}.}


\bibitem[\protect\citeauthoryear{IBM}{IBM}{2010}]%
        {IBM2010}
{IBM}. 2010.
\newblock Capitalizing on Complexity.
\newblock   (2010).
\newblock
\newblock
\shownote{Available at
  \url{http://www-935.ibm.com/services/us/ceo/ceostudy2010/multimedia.html}.}


\bibitem[\protect\citeauthoryear{IBM}{IBM}{2016a}]%
        {IBMbd}
{IBM}. 2016a.
\newblock {IBM} Analytics and Big Data.
\newblock   (2016).
\newblock
\newblock
\shownote{Available at \url{http://www.ibm.com/analytics/us/en/ or
  http://www-01.ibm.com/software/data/bigdata/}.}


\bibitem[\protect\citeauthoryear{IBM}{IBM}{2016b}]%
        {ibmds}
{IBM}. 2016b.
\newblock What is a data scientist?
\newblock   (2016).
\newblock
\newblock
\shownote{Available at
  \url{http://www-01.ibm.com/software/data/infosphere/data-scientist/}.}


\bibitem[\protect\citeauthoryear{IDA}{IDA}{2014}]%
        {IDA}
{IDA}. 2014.
\newblock International Institute of Data \& Analytics.
\newblock   (2014).
\newblock
\newblock
\shownote{Available at \url{www.datasciences.org}.}


\bibitem[\protect\citeauthoryear{IEEEBD}{IEEEBD}{2014}]%
        {IEEEBD}
{IEEEBD}. 2014.
\newblock {IEEE} Big Data Initiative.
\newblock   (2014).
\newblock
\newblock
\shownote{Available at \url{http://bigdata.ieee.org/}.}


\bibitem[\protect\citeauthoryear{IFSC-96}{IFSC-96}{1996}]%
        {IFSC-96}
{IFSC-96}. 1996.
\newblock \showarticletitle{Data Science, Classification, and Related Methods}.
\newblock  (1996).
\newblock
\newblock
\shownote{Available at \url{http://d-nb.info/955715512/04}.}


\bibitem[\protect\citeauthoryear{IJDS}{IJDS}{2016}]%
        {IJDS}
{IJDS}. 2016.
\newblock International Journal of Data Science.
\newblock   (2016).
\newblock
\newblock
\shownote{Available at \url{http://www.inderscience.com/jhome.php?jcode=ijds}.}


\bibitem[\protect\citeauthoryear{IJRDS}{IJRDS}{2017}]%
        {IJRDS}
{IJRDS}. 2017.
\newblock International Journal of Research on Data Science.
\newblock   (2017).
\newblock
\newblock
\shownote{Available at
  \url{http://www.sciencepublishinggroup.com/journal/index?journalid=310}.}


\bibitem[\protect\citeauthoryear{INFORMS}{INFORMS}{2014}]%
        {INFORMS14c}
{INFORMS}. 2014.
\newblock Candidate Handbook.
\newblock   (2014).
\newblock
\newblock
\shownote{Available at
  \url{https://www.informs.org/Certification-Continuing-Ed/Analytics-Certification/Candidate-Handbook}.}


\bibitem[\protect\citeauthoryear{INFORMS}{INFORMS}{2016}]%
        {INFORMS}
{INFORMS}. 2016.
\newblock Institute for Operations Research and the Management Sciences.
\newblock   (2016).
\newblock
\newblock
\shownote{Available at \url{https://www.informs.org/}.}


\bibitem[\protect\citeauthoryear{Iwata}{Iwata}{2008}]%
        {Iwata08}
{S Iwata}. 2008.
\newblock \showarticletitle{Scientific {``agenda''} of data science}.
\newblock {\em Data Science Journal\/} {7}, 5 (2008), 54--56.
\newblock


\bibitem[\protect\citeauthoryear{Jagadish, Gehrke, Labrinidis,
  Papakonstantinou, Patel, Ramakrishnan, and Shahabi}{Jagadish
  et~al\mbox{.}}{2014}]%
        {Jagadish14}
{H.V. Jagadish}, {Johannes Gehrke}, {Alexandros Labrinidis}, {Yannis
  Papakonstantinou}, {Jignesh~M. Patel}, {Raghu Ramakrishnan}, {and} {Cyrus
  Shahabi}. 2014.
\newblock \showarticletitle{Big data and its technical challenges}.
\newblock {\it Commun. ACM} {57}, 7 (2014), 86--94.
\newblock


\bibitem[\protect\citeauthoryear{Jagadish}{Jagadish}{2015}]%
        {Jagadish15}
{H.~V. Jagadish}. 2015.
\newblock \showarticletitle{Big data and science: Myths and reality}.
\newblock {\em Big Data Research\/} {2}, 2 (2015), 49--52.
\newblock


\bibitem[\protect\citeauthoryear{JDS}{JDS}{2002}]%
        {JDS}
{JDS}. 2002.
\newblock Journal of Data Science.
\newblock   (2002).
\newblock
\newblock
\shownote{Available at \url{http://www.jds-online.com/}.}


\bibitem[\protect\citeauthoryear{JDSA}{JDSA}{2015}]%
        {JDSA}
{JDSA}. 2015.
\newblock International Journal of Data Science and Analytics {(JDSA)}.
\newblock   (2015).
\newblock
\newblock
\shownote{Available at \url{http://www.springer.com/41060}.}


\bibitem[\protect\citeauthoryear{JFDS}{JFDS}{2016}]%
        {JFDS}
{JFDS}. 2016.
\newblock The Journal of Finance and Data Science.
\newblock   (2016).
\newblock
\newblock
\shownote{Available at
  \url{http://www.keaipublishing.com/en/journals/the-journal-of-finance-and-data-science/}.}


\bibitem[\protect\citeauthoryear{Kaggle}{Kaggle}{2016}]%
        {Kaggledata}
{Kaggle}. 2016.
\newblock Kaggle Competition Data.
\newblock   (2016).
\newblock
\newblock
\shownote{Available at \url{https://www.kaggle.com/competitions}.}


\bibitem[\protect\citeauthoryear{Kalidindi}{Kalidindi}{2015}]%
        {Kalidindi15-2}
{Surya~R. Kalidindi}. 2015.
\newblock \showarticletitle{Data science and cyberinfrastructure: critical
  enablers for accelerated development of hierarchical materials}.
\newblock {\em International Materials Reviews\/} {60}, 3 (2015), 150--168.
\newblock


\bibitem[\protect\citeauthoryear{KDD89}{KDD89}{1989}]%
        {KDD89}
{KDD89}. 1989.
\newblock {IJCAI-89} Workshop on Knowledge Discovery in Databases.
\newblock   (1989).
\newblock
\newblock
\shownote{Available at
  \url{http://www.kdnuggets.com/meetings/kdd89/index.html}.}


\bibitem[\protect\citeauthoryear{KDnuggets}{KDnuggets}{2015}]%
        {kdnuggets-vis}
{KDnuggets}. 2015.
\newblock Visualization Software.
\newblock   (2015).
\newblock
\newblock
\shownote{Available at
  \url{http://www.kdnuggets.com/software/visualization.html}.}


\bibitem[\protect\citeauthoryear{Kdnuggets}{Kdnuggets}{2016}]%
        {kdnuggets}
{Kdnuggets}. 2016.
\newblock Kdnuggets.
\newblock   (2016).
\newblock
\newblock
\shownote{Available at \url{http://www.kdnuggets.com/}.}


\bibitem[\protect\citeauthoryear{Kelly}{Kelly}{2012}]%
        {Kelly12}
{K Kelly}. 2012.
\newblock \showarticletitle{The Quantified Century}. In {\em Quantified Self
  Conference}.
\newblock
\newblock
\shownote{Available at
  \url{http://quantifiedself.com/conference/Palo-Alto-2012}.}


\bibitem[\protect\citeauthoryear{Khan, Yaqoob, Hashem, and et~al}{Khan
  et~al\mbox{.}}{2014}]%
        {Khan14}
{Nawsher Khan}, {Ibrar Yaqoob}, {Ibrahim Abaker~Targio Hashem}, {and} {et al}.
  2014.
\newblock \showarticletitle{Big data: Survey, technologies, opportunities, and
  challenges}.
\newblock {\em The Scientific World Journal\/}  {2014} (2014), 18.
\newblock


\bibitem[\protect\citeauthoryear{King and Magoulas}{King and Magoulas}{2015}]%
        {King15}
{John King} {and} {Roger Magoulas}. 2015.
\newblock 2015 Data Science Salary Survey.
\newblock   (2015).
\newblock
\newblock
\shownote{Available at
  \url{http://duu86o6n09pv.cloudfront.net/reports/2015-data-science-salary-survey.pdf}.}


\bibitem[\protect\citeauthoryear{Kohavi, Rothleder, and Simoudis}{Kohavi
  et~al\mbox{.}}{2002}]%
        {Kohavi02}
{Ron Kohavi}, {Neal~J. Rothleder}, {and} {Evangelos Simoudis}. 2002.
\newblock \showarticletitle{Emerging trends in business analytics}.
\newblock {\it Commun. ACM} {45}, 8 (2002), 45--48.
\newblock


\bibitem[\protect\citeauthoryear{Lab}{Lab}{2016}]%
        {MLBase}
{AMP Lab}. 2016.
\newblock MLBase.
\newblock   (2016).
\newblock
\newblock
\shownote{Available at \url{http://mlbase.org/}.}


\bibitem[\protect\citeauthoryear{Labrinidis and Jagadish}{Labrinidis and
  Jagadish}{2012}]%
        {Labrinidis12}
{A. Labrinidis} {and} {H.~V. Jagadish}. 2012.
\newblock \showarticletitle{Challenges and opportunities with Big Data}.
\newblock {\em Proceedings of the VLDB Endowment\/} {5}, 12 (2012), 2032--2033.
\newblock


\bibitem[\protect\citeauthoryear{Laney}{Laney}{2001}]%
        {Laney01}
{Douglas Laney}. 2001.
\newblock {3D} Data Management: Controlling Data Volume, Velocity and Variety.
\newblock   (2001).
\newblock
\newblock
\shownote{Technical Report, META Group.}


\bibitem[\protect\citeauthoryear{Lazer, Kennedy, King, and Vespignani}{Lazer
  et~al\mbox{.}}{2014}]%
        {Lazer14}
{LD. Lazer}, {R Kennedy}, {G King}, {and} {A Vespignani}. 2014.
\newblock \showarticletitle{The parable of Google flu: Traps in big data
  analysis}.
\newblock {\em Science\/}  {343} (2014), 1203--1205.
\newblock


\bibitem[\protect\citeauthoryear{LDC}{LDC}{2016}]%
        {LDC15}
{LDC}. 2016.
\newblock Linguistic Data Consortium.
\newblock   (2016).
\newblock
\newblock
\shownote{Available at \url{https://www.ldc.upenn.edu/about}.}


\bibitem[\protect\citeauthoryear{LinkedIn}{LinkedIn}{2016}]%
        {linkedinds}
{LinkedIn}. 2016.
\newblock LinkedIn Jobs.
\newblock   (2016).
\newblock
\newblock
\shownote{Available at
  \url{https://www.linkedin.com/jobs/data-scientist-jobs}.}


\bibitem[\protect\citeauthoryear{Loukides}{Loukides}{2011}]%
        {Loukides11}
{Mike Loukides}. 2011.
\newblock {\em The Evolution of Data Products}.
\newblock O'Reilly, Cambridge.
\newblock


\bibitem[\protect\citeauthoryear{Loukides}{Loukides}{2012}]%
        {Loukides12}
{Mike Loukides}. 2012.
\newblock {\em What is data science?}
\newblock O'Reilly Media, Sebastopol, CA.
\newblock


\bibitem[\protect\citeauthoryear{Manieri, Brewer, Riestra, Demchenko, Hemmje,
  Wiktorski, Ferrari, and Frey}{Manieri et~al\mbox{.}}{2015}]%
        {Manieri15-1}
{Andrea Manieri}, {Steve Brewer}, {Ruben Riestra}, {Yuri Demchenko}, {Matthias
  Hemmje}, {Tomasz Wiktorski}, {Tiziana Ferrari}, {and} {Jérémy Frey}. 2015.
\newblock \showarticletitle{Data science professional uncovered: How the
  {EDISON} project will contribute to a widely accepted profile for data
  scientists}. In {\em 2015 IEEE 7th International Conference on Cloud
  Computing Technology and Science (CloudCom)}. 588--593.
\newblock


\bibitem[\protect\citeauthoryear{Matsudaira}{Matsudaira}{2015}]%
        {Matsudaira15}
{Kate Matsudaira}. 2015.
\newblock \showarticletitle{The science of managing data science}.
\newblock {\it Commun. ACM} {58}, 6 (2015), 44--47.
\newblock


\bibitem[\protect\citeauthoryear{McKinsey}{McKinsey}{2011}]%
        {Mckinsey11}
{McKinsey}. 2011.
\newblock Big Data: The Next Frontier for Innovation, CCompetition, and
  Productivity.
\newblock   (2011).
\newblock
\newblock
\shownote{McKinsey Global Institute.}


\bibitem[\protect\citeauthoryear{Miller}{Miller}{2013}]%
        {Miller13}
{Claire~Cain Miller}. 2013.
\newblock \showarticletitle{Data Science: The Numbers of Our Lives}.
\newblock {\em New York Times\/} (2013).
\newblock


\bibitem[\protect\citeauthoryear{Morrell}{Morrell}{1968}]%
        {Morrell68}
{AJH. Morrell}. 1968.
\newblock \showarticletitle{Information processing 68 (Ed.)}. In {\em
  Proceedings of IFIP Congress 1968}. Edinburgh, UK.
\newblock


\bibitem[\protect\citeauthoryear{Murray-Rust}{Murray-Rust}{2007}]%
        {Rust07}
{Peter Murray-Rust}. 2007.
\newblock \showarticletitle{{Data-Driven} Science: A Scientist's View}. In {\em
  NSF/JISC 2007 Digital Repositories Workshop}.
\newblock
\newblock
\shownote{Available at
  \url{http://www.sis.pitt.edu/repwkshop/papers/murray.pdf}.}


\bibitem[\protect\citeauthoryear{Naur}{Naur}{1968}]%
        {Naur68}
{Peter Naur}. 1968.
\newblock \showarticletitle{`Datalogy', the science of data and data
  processes}.
\newblock  (1968), 1383--1387.
\newblock


\bibitem[\protect\citeauthoryear{Naur}{Naur}{1974}]%
        {Naur74}
{Peter Naur}. 1974.
\newblock {\em Concise Survey of Computer Methods}.
\newblock Studentlitteratur, Lund, Sweden.
\newblock
\showISBNx{91-44-07881-1}


\bibitem[\protect\citeauthoryear{NCSU}{NCSU}{2007a}]%
        {IAA}
{NCSU}. 2007a.
\newblock Institute for Advanced Analytics, North Carolina State University.
\newblock   (2007).
\newblock
\newblock
\shownote{Available at \url{http://analytics.ncsu.edu/}.}


\bibitem[\protect\citeauthoryear{NCSU}{NCSU}{2007b}]%
        {MAS07}
{NCSU}. 2007b.
\newblock Master of Science in Analytics, Institute for Advanced Analytics,
  North Carolina State University.
\newblock   (2007).
\newblock
\newblock
\shownote{Available at \url{http://analytics.ncsu.edu/}.}


\bibitem[\protect\citeauthoryear{Nelson}{Nelson}{2009}]%
        {Nelson09}
{Michael~L. Nelson}. 2009.
\newblock \showarticletitle{{Data-driven} science: A new paradigm?}
\newblock {\em EDUCAUSE Review\/} {44}, 4 (2009), 6--7.
\newblock


\bibitem[\protect\citeauthoryear{NICTA}{NICTA}{2016}]%
        {NICTA}
{NICTA}. 2016.
\newblock National ICT Australia.
\newblock   (2016).
\newblock
\newblock
\shownote{Available at \url{https://www.nicta.com.au/}.}


\bibitem[\protect\citeauthoryear{NIST}{NIST}{2015}]%
        {Trec15}
{NIST}. 2015.
\newblock {NIST} Text Retrieval Conference Data.
\newblock   (2015).
\newblock
\newblock
\shownote{Available at \url{http://trec.nist.gov/data.html}.}


\bibitem[\protect\citeauthoryear{NSB}{NSB}{2005}]%
        {NSB05}
{NSB}. 2005.
\newblock \showarticletitle{Long-lived Digital Data Collections: Enabling
  Research and Education in the 21st Century}.
\newblock  (2005).
\newblock


\bibitem[\protect\citeauthoryear{NSF}{NSF}{2007}]%
        {NSF0728}
{NSF}. 2007.
\newblock \showarticletitle{{US} {NSF07-28}}.
\newblock  (2007).
\newblock
\newblock
\shownote{Available at \url{http://www.nsf.gov/pubs/2007/nsf0728/nsf0728.pdf}.}


\bibitem[\protect\citeauthoryear{OECD}{OECD}{2007}]%
        {OECD04}
{OECD}. 2007.
\newblock {OECD} Principles and Guidances for Access to Research Data from
  Public Funding.
\newblock   (2007).
\newblock
\newblock
\shownote{Available at \url{https://www.oecd.org/sti/sci-tech/38500813.pdf}.}


\bibitem[\protect\citeauthoryear{OPENedX}{OPENedX}{2016}]%
        {OPENedX}
{OPENedX}. 2016.
\newblock {OPENedX} Online education platform.
\newblock   (2016).
\newblock
\newblock
\shownote{Available at \url{https://open.edx.org/}.}


\bibitem[\protect\citeauthoryear{O'Reilly}{O'Reilly}{2005}]%
        {O'reilly05}
{Tim O'Reilly}. 2005.
\newblock What is Web 2.0.
\newblock   (2005).
\newblock
\newblock
\shownote{Available at
  \url{http://oreilly.com/pub/a/web2/archive/what-is-web-20.html?page=3}.}


\bibitem[\protect\citeauthoryear{Patil}{Patil}{2011}]%
        {Patil11}
{DJ Patil}. 2011.
\newblock {\em Building Data Science Teams}.
\newblock O'Reilly Media.
\newblock


\bibitem[\protect\citeauthoryear{Paulk, Curtis, Chrissis, and Weber}{Paulk
  et~al\mbox{.}}{1993}]%
        {Paulk93}
{M.~C. Paulk}, {B. Curtis}, {M.~B. Chrissis}, {and} {C. Weber}. 1993.
\newblock \showarticletitle{Capability maturity model Version 1.1}.
\newblock {\em IEEE Software\/} {10}, 4 (1993), 18--27.
\newblock


\bibitem[\protect\citeauthoryear{Press}{Press}{2013}]%
        {Press13}
{Gil Press}. 2013.
\newblock A Very Short History Of Data Science.
\newblock   (2013).
\newblock
\newblock
\shownote{Available at
  \url{http://www.forbes.com/sites/gilpress/2013/05/28/a-very-short-history-of-data-science/#61ae3ebb69fd}.}


\bibitem[\protect\citeauthoryear{Qian}{Qian}{1991}]%
        {Qian91}
{Xuesen Qian}. 1991.
\newblock \showarticletitle{Revisiting issues on open complex giant systems}.
\newblock {\em Pattern Recognit. Artif. Intell.\/} {4}, 1 (1991), 5--8.
\newblock


\bibitem[\protect\citeauthoryear{Qian, Yu, and Dai}{Qian et~al\mbox{.}}{1993}]%
        {Qian93}
{Xuesen Qian}, {Jingyuan Yu}, {and} {Ruwei Dai}. 1993.
\newblock \showarticletitle{A new discipline of science-The study of open
  complex giant system and its methodology}.
\newblock {\em Chin. J. Syst. Eng. Electron.\/} {4}, 2 (1993), 2--12.
\newblock


\bibitem[\protect\citeauthoryear{RapidMiner}{RapidMiner}{2016}]%
        {Rapidminer}
{RapidMiner}. 2016.
\newblock RapidMiner.
\newblock   (2016).
\newblock
\newblock
\shownote{Available at \url{https://rapidminer.com/}.}


\bibitem[\protect\citeauthoryear{Renae}{Renae}{2011}]%
        {Vance11}
{Samantha Renae}. 2011.
\newblock \showarticletitle{Data analytics: Crunching the future}.
\newblock {\em Bloomberg Businessweek\/} (2011).
\newblock
\newblock
\shownote{September 8.}


\bibitem[\protect\citeauthoryear{Review}{Review}{2016}]%
        {solutionsreview-datainteg}
{Solutions Review}. 2016.
\newblock Data Integration and Application Integration Solutions Directory.
\newblock   (2016).
\newblock
\newblock
\shownote{Available at
  \url{http://solutionsreview.com/data-integration/data-integration-solutions-directory/}.}


\bibitem[\protect\citeauthoryear{Rudin}{Rudin}{2014}]%
        {Rudin14}
{C. Rudin}. 2014.
\newblock Discovery with Data: Leveraging Statistics with Computer Science to
  Transform Science and Society.
\newblock   (2014).
\newblock
\newblock
\shownote{American Statistical Association.}


\bibitem[\protect\citeauthoryear{SAS}{SAS}{2013}]%
        {SAS13}
{SAS}. 2013.
\newblock Big Data Analytics: An Assessment of Demand for Labour and Dkills,
  2012-2017.
\newblock   (2013).
\newblock
\newblock
\shownote{Available at
  \url{https://www.thetechpartnership.com/globalassets/pdfs/research-2014/bigdata_report_nov14.pdf}.}


\bibitem[\protect\citeauthoryear{SAS}{SAS}{2016}]%
        {SAS16}
{SAS}. 2016.
\newblock {SAS} Insights.
\newblock   (2016).
\newblock
\newblock
\shownote{Available at \url{http://www.sas.com/en_us/insights.html}.}


\bibitem[\protect\citeauthoryear{Schoenherr and Speier-Pero}{Schoenherr and
  Speier-Pero}{2015}]%
        {Schoenherr-jbl13}
{Tobias Schoenherr} {and} {Cheri Speier-Pero}. 2015.
\newblock \showarticletitle{Data science, predictive analytics, and big data in
  supply chain management: Current state and future potential}.
\newblock {\em Journal of Business Logistics\/} {36}, 1 (2015), 120--132.
\newblock


\bibitem[\protect\citeauthoryear{Security}{Security}{2015}]%
        {CIS2015}
{China~Information Security}. 2015.
\newblock Big Data Strategies and Actions in Major Countries.
\newblock   (2015).
\newblock
\newblock
\shownote{Available at
  \url{http://www.cac.gov.cn/2015-07/03/c_1115812491.htm}.}


\bibitem[\protect\citeauthoryear{SIAM}{SIAM}{2016}]%
        {siamjobs}
{SIAM}. 2016.
\newblock SIAM career center.
\newblock   (2016).
\newblock
\newblock
\shownote{Available at \url{http://jobs.siam.org/home/}.}


\bibitem[\protect\citeauthoryear{Siart, Kopp, and Apel}{Siart
  et~al\mbox{.}}{2015}]%
        {Siart15}
{C. Siart}, {S. Kopp}, {and} {J. Apel}. 2015.
\newblock \showarticletitle{The interface between data science, research
  assessment and science support - Highlights from the German perspective and
  examples from Heidelberg University}. In {\em 2015 IIAI 4th International
  Congress on Advanced Applied Informatics (IIAI-AAI)}. 472--476.
\newblock


\bibitem[\protect\citeauthoryear{Silk}{Silk}{2016}]%
        {Silk}
{Silk}. 2016.
\newblock Data Science University Programs.
\newblock   (2016).
\newblock
\newblock
\shownote{Available at \url{http://data-science-university-programs.silk.co/}.}


\bibitem[\protect\citeauthoryear{Smarr}{Smarr}{2012}]%
        {Smarr12}
{Larry Smarr}. 2012.
\newblock \showarticletitle{Quantifying your body: A how-to guide from a
  systems biology perspective}.
\newblock {\em Biotechnology Journal\/} {7}, 8 (2012), 980--991.
\newblock
\showISSN{1860-7314}
\showDOI{%
\url{http://dx.doi.org/10.1002/biot.201100495}}


\bibitem[\protect\citeauthoryear{Smith}{Smith}{2006}]%
        {Smith06}
{F.~Jack Smith}. 2006.
\newblock \showarticletitle{Data science as an academic discipline}.
\newblock {\em Data Science Journal\/}  {5} (2006), 163--164.
\newblock


\bibitem[\protect\citeauthoryear{SSDS}{SSDS}{2015}]%
        {SSDS}
{SSDS}. 2015.
\newblock Springer Series in the Data Sciences.
\newblock   (2015).
\newblock
\newblock
\shownote{Available at \url{http://www.springer.com/series/13852}.}


\bibitem[\protect\citeauthoryear{Stanford}{Stanford}{2014}]%
        {Stanford}
{Stanford}. 2014.
\newblock Stanford Data Science Initiatives, Stanford University.
\newblock   (2014).
\newblock
\newblock
\shownote{Available at \url{https://sdsi.stanford.edu/}.}


\bibitem[\protect\citeauthoryear{Stewart and McMillan}{Stewart and
  McMillan}{1987}]%
        {Stewart87}
{Thomas~R. Stewart} {and} {Jr.~Claude McMillan}. 1987.
\newblock \showarticletitle{Descriptive and prescriptive models for judgment
  and decision making: Implications for knowledge engineering}. In {\em Expert
  Judgment and Expert Systems}, {Jeryl~L. Mumpower}, {Ortwin Renn},
  {Lawrence~D. Phillips}, {and} {V.~R. R.~Uppuluri (Eds.)} (Eds.).
  Springer-Verlag, London, 305--320.
\newblock


\bibitem[\protect\citeauthoryear{Stonebraker, Madden, and Dubey}{Stonebraker
  et~al\mbox{.}}{2013}]%
        {Stonebraker13}
{Michael Stonebraker}, {Sam Madden}, {and} {Pradeep Dubey}. 2013.
\newblock \showarticletitle{Intel `big data' science and technology center
  vision and execution plan}.
\newblock {\em SIGMOD Record\/} {42}, 1 (2013), 44--49.
\newblock


\bibitem[\protect\citeauthoryear{Swan and Brown}{Swan and Brown}{2008}]%
        {jisc08}
{Alma Swan} {and} {Sheridan Brown}. 2008.
\newblock \showarticletitle{The Skills, Role \& Career Structure of Data
  Scientists \& Curators: Assessment of Current Practice \& Future Needs}.
\newblock  (2008).
\newblock
\newblock
\shownote{Technical Report. University of Southampton.}


\bibitem[\protect\citeauthoryear{Swan}{Swan}{2013}]%
        {Swan13}
{Melanie Swan}. 2013.
\newblock \showarticletitle{The quantified self: Fundamental disruption in big
  data science and biological discovery}.
\newblock {\em Big Data\/} {1}, 2 (2013), 85--99.
\newblock


\bibitem[\protect\citeauthoryear{Technavio}{Technavio}{2016}]%
        {Technavio-healthana}
{Technavio}. 2016.
\newblock Top 10 Healthcare Data Analytics Companies.
\newblock   (2016).
\newblock
\newblock
\shownote{Available at
  \url{http://www.technavio.com/blog/top-10-healthcare-data-analytics-companies}.}


\bibitem[\protect\citeauthoryear{TFDSAA}{TFDSAA}{2013}]%
        {TFDSAA}
{TFDSAA}. 2013.
\newblock {IEEE} Task Force on Data Science and Advanced Analytics.
\newblock   (2013).
\newblock
\newblock
\shownote{Available at \url{http://dsaatf.dsaa.co/}.}


\bibitem[\protect\citeauthoryear{TOBD}{TOBD}{2015}]%
        {TOBD}
{TOBD}. 2015.
\newblock {IEEE} Transactions on Big Data.
\newblock   (2015).
\newblock
\newblock
\shownote{Available at \url{https://www.computer.org/web/tbd}.}


\bibitem[\protect\citeauthoryear{Today}{Today}{2016}]%
        {predictiveanalyticstoday-dataprep}
{Predictive~Analytics Today}. 2016.
\newblock 29 Data Preparation Tools and Platforms.
\newblock   (2016).
\newblock
\newblock
\shownote{Available at
  \url{http://www.predictiveanalyticstoday.com/data-preparation-tools-and-platforms/}.}


\bibitem[\protect\citeauthoryear{Tukey}{Tukey}{1962}]%
        {Tukey62}
{John~W. Tukey}. 1962.
\newblock \showarticletitle{The future of data analysis}.
\newblock {\em Ann. Math. Statist.\/} {33}, 1 (1962), 1--67.
\newblock


\bibitem[\protect\citeauthoryear{Tukey}{Tukey}{1977}]%
        {Tukey77}
{John~W. Tukey}. 1977.
\newblock {\em Exploratory Data Analysis}.
\newblock Pearson.
\newblock
\showISBNx{978-0201076165}


\bibitem[\protect\citeauthoryear{Tutiempo}{Tutiempo}{2016}]%
        {Climatedata}
{Tutiempo}. 2016.
\newblock Global Climate Data.
\newblock   (2016).
\newblock
\newblock
\shownote{Available at \url{http://en.tutiempo.net/climate}.}


\bibitem[\protect\citeauthoryear{UCI}{UCI}{2016}]%
        {UCIdata}
{UCI}. 2016.
\newblock {UCI} Machine Learning Repository.
\newblock   (2016).
\newblock
\newblock
\shownote{Available at \url{archive.ics.uci.edu/ml/}.}


\bibitem[\protect\citeauthoryear{Udacity}{Udacity}{2016}]%
        {Udacity}
{Udacity}. 2016.
\newblock Udacity Courses.
\newblock   (2016).
\newblock
\newblock
\shownote{Available at \url{https://www.udacity.com/courses/data-science}.}


\bibitem[\protect\citeauthoryear{Udemy}{Udemy}{2016}]%
        {Udemy}
{Udemy}. 2016.
\newblock Udemy Courses.
\newblock   (2016).
\newblock
\newblock
\shownote{Available at
  \url{https://www.udemy.com/courses/search/?ref=home&src=ukw&q=data+science&lang=en}.}


\bibitem[\protect\citeauthoryear{UK}{UK}{2016}]%
        {UKbd}
{UK}. 2016.
\newblock {UK} Big Data.
\newblock   (2016).
\newblock
\newblock
\shownote{Available at
  \url{http://www.rcuk.ac.uk/research/infrastructure/big-data/}.}


\bibitem[\protect\citeauthoryear{UK-HM}{UK-HM}{2012}]%
        {UKpd}
{UK-HM}. 2012.
\newblock \showarticletitle{UK HM Government}.
\newblock  (2012).
\newblock
\newblock
\shownote{Available at
  \url{http://data.gov.uk/sites/default/files/Open_data_White_Paper.pdf}.}


\bibitem[\protect\citeauthoryear{UK-OD}{UK-OD}{2016}]%
        {UKod}
{UK-OD}. 2016.
\newblock {UK} Open Data.
\newblock   (2016).
\newblock
\newblock
\shownote{Available at \url{http://data.gov.uk/}.}


\bibitem[\protect\citeauthoryear{UMichi}{UMichi}{2015}]%
        {MIDS}
{UMichi}. 2015.
\newblock Michigan Institute For Data Science, University of Michigan.
\newblock   (2015).
\newblock
\newblock
\shownote{Available at \url{http://midas.umich.edu/}.}


\bibitem[\protect\citeauthoryear{UN}{UN}{2010}]%
        {UNpulse}
{UN}. 2010.
\newblock United Nation Global Pulse Projects.
\newblock   (2010).
\newblock
\newblock
\shownote{Available at \url{http://www.unglobalpulse.org/}.}


\bibitem[\protect\citeauthoryear{US-OD}{US-OD}{2016}]%
        {usod}
{US-OD}. 2016.
\newblock {US} government open data.
\newblock   (2016).
\newblock
\newblock
\shownote{Available at \url{https://www.data.gov/}.}


\bibitem[\protect\citeauthoryear{USD2D}{USD2D}{2016}]%
        {USd2d}
{USD2D}. 2016.
\newblock {US} National Consortium for Data Science.
\newblock   (2016).
\newblock
\newblock
\shownote{Available at \url{data2discovery.org}.}


\bibitem[\protect\citeauthoryear{USDSC}{USDSC}{2016}]%
        {USDSC}
{USDSC}. 2016.
\newblock {US} Degree Programs in Analytics and Data Science.
\newblock   (2016).
\newblock
\newblock
\shownote{Available at \url{http://analytics.ncsu.edu/?page_id=4184}.}


\bibitem[\protect\citeauthoryear{USNSF}{USNSF}{2012}]%
        {USNSF}
{USNSF}. 2012.
\newblock {US} Big Data Research Initiative.
\newblock   (2012).
\newblock
\newblock
\shownote{Available at \url{http://www.nsf.gov/cise/news/bigdata.jsp}.}


\bibitem[\protect\citeauthoryear{UTS}{UTS}{2011}]%
        {UTSA}
{UTS}. 2011.
\newblock Master of Analytics (Research) and Doctor of Philosophy Thesis:
  Analytics, {{Advanced} Analytics Institute, University of Technology Sydney}.
\newblock   (2011).
\newblock
\newblock
\shownote{Available at
  \url{http://www.uts.edu.au/research-and-teaching/our-research/advanced-analytics-institute/education-and-research-opportuniti-1}.}


\bibitem[\protect\citeauthoryear{UTSAAI}{UTSAAI}{2011}]%
        {UTSAAI}
{UTSAAI}. 2011.
\newblock Advanced Analytics Institute, University of Technology Sydney.
\newblock   (2011).
\newblock
\newblock
\shownote{Available at \url{https://analytics.uts.edu.au/}.}


\bibitem[\protect\citeauthoryear{van Dyk, Fuentes, Jordan, Newton, Ray, Lang,
  and Wickham}{van Dyk et~al\mbox{.}}{2015}]%
        {amstatnews15}
{David van Dyk}, {Montse Fuentes}, {Michael~I. Jordan}, {Michael Newton},
  {Bonnie~K. Ray}, {Duncan~Temple Lang}, {and} {Hadley Wickham}. 2015.
\newblock {ASA} Statement on the Role of Statistics in Data Science.
\newblock   (2015).
\newblock
\newblock
\shownote{Available at
  \url{http://magazine.amstat.org/blog/2015/10/01/asa-statement-on-the-role-of-statistics-in-data-science/}.}


\bibitem[\protect\citeauthoryear{Vast}{Vast}{2016}]%
        {Vast15}
{Vast}. 2016.
\newblock Visual Analytics Community.
\newblock   (2016).
\newblock
\newblock
\shownote{Available at \url{http://vacommunity.org/HomePage}.}


\bibitem[\protect\citeauthoryear{Vesset, Woo, Morris, Villars, Little, Bozman,
  Borovick, Olofson, Feldman, Conway, Eastwood, and Yezhkova}{Vesset
  et~al\mbox{.}}{2012}]%
        {IDC12}
{Dan Vesset}, {Benjamin Woo}, {Henry~D. Morris}, {Richard~L. Villars}, {Gard
  Little}, {Jean~S. Bozman}, {Lucinda Borovick}, {Carl~W. Olofson}, {Susan
  Feldman}, {Steve Conway}, {Matthew Eastwood}, {and} {Natalya Yezhkova}. 2012.
\newblock {IDC} Worldwide Big Data Technology and Services 2012-2015 Forecast.
\newblock   (2012).
\newblock


\bibitem[\protect\citeauthoryear{Viseu and Suchman}{Viseu and Suchman}{2010}]%
        {Viseu10}
{Ana Viseu} {and} {Lucy Suchman}. 2010.
\newblock {\em Wearable Augmentations: Imaginaries of the Informed Body}.
\newblock Berghahn Books, New York, NY, USA, 161--184.
\newblock
\showISBNx{9781845456641}


\bibitem[\protect\citeauthoryear{Whitehouse}{Whitehouse}{2015}]%
        {whcds}
{Whitehouse}. 2015.
\newblock The White House Names Dr. {DJ} Patil as the First {U.S.} Chief Data
  Scientist.
\newblock   (2015).
\newblock
\newblock
\shownote{Available at
  \url{https://www.whitehouse.gov/blog/2015/02/18/white-house-names-dr-dj-patil-first-us-chief-data-scientist}.}


\bibitem[\protect\citeauthoryear{Wikipedia}{Wikipedia}{2016a}]%
        {wikipedia-hpc}
{Wikipedia}. 2016a.
\newblock Comparison of cluster software.
\newblock   (2016).
\newblock
\newblock
\shownote{Available at
  \url{https://en.wikipedia.org/wiki/Comparison_of_cluster_software}.}


\bibitem[\protect\citeauthoryear{Wikipedia}{Wikipedia}{2016b}]%
        {Informatics}
{Wikipedia}. 2016b.
\newblock Informatics.
\newblock   (2016).
\newblock
\newblock
\shownote{Available at \url{https://en.wikipedia.org/wiki/Informatics}.}


\bibitem[\protect\citeauthoryear{Wikipedia}{Wikipedia}{2016c}]%
        {Wikipedia-reporting}
{Wikipedia}. 2016c.
\newblock List of reporting software.
\newblock   (2016).
\newblock
\newblock
\shownote{Available at
  \url{https://en.wikipedia.org/wiki/List_of_reporting_software}.}


\bibitem[\protect\citeauthoryear{WIRED}{WIRED}{2014}]%
        {wired14}
{WIRED}. 2014.
\newblock How Europe can seize the starring role in big data.
\newblock   (2014).
\newblock
\newblock
\shownote{Available at \url{www.wired.com/insights/2014/09/europe-big-data/}.}


\bibitem[\protect\citeauthoryear{Wolf}{Wolf}{2012}]%
        {Wolf12}
{G. Wolf}. 2012.
\newblock \showarticletitle{The {Data-Driven} Life}.
\newblock {\em New York Times\/} (2012).
\newblock


\bibitem[\protect\citeauthoryear{Wu}{Wu}{1997}]%
        {wu97}
{Jeff Wu}. 1997.
\newblock Statistics = Data Science?
\newblock   (1997).
\newblock
\newblock
\shownote{Available at
  \url{http://www2.isye.gatech.edu/~jeffwu/presentations/datascience.pdf}.}


\bibitem[\protect\citeauthoryear{Yahoo}{Yahoo}{2016}]%
        {Yahoofinance}
{Yahoo}. 2016.
\newblock Yahoo Finance.
\newblock   (2016).
\newblock
\newblock
\shownote{Available at \url{finance.yahoo.com}.}


\bibitem[\protect\citeauthoryear{Yau}{Yau}{2009}]%
        {Yau09}
{Nathan Yau}. 2009.
\newblock Rise of the Data Scientist.
\newblock   (2009).
\newblock
\newblock
\shownote{Available at
  \url{http://flowingdata.com/2009/06/04/rise-of-the-data-scientist/}.}


\bibitem[\protect\citeauthoryear{Yiu}{Yiu}{2012}]%
        {yiu12}
{Chris Yiu}. 2012.
\newblock The Big Data Opportunity.
\newblock   (2012).
\newblock
\newblock
\shownote{Available at
  \url{http://www.policyexchange.org.uk/images/publications/the\%20big\%20data\%20opportunity.pdf}.}


\bibitem[\protect\citeauthoryear{Yu}{Yu}{2014}]%
        {yu14}
{Bin Yu}. 2014.
\newblock \showarticletitle{{IMS} Presidential Address: Let us own Data
  Science}.
\newblock {\em IMS Bulletin Online\/} (2014).
\newblock
\newblock
\shownote{1 Oct 2014.}


\end{thebibliography}

%
%
%
%
%

\received{February 2007}{March 2009}{June 2009}



%
%
%
%

\end{document}